\documentclass[%
 reprint,
superscriptaddress,
groupedaddress,
 amsmath,amssymb,showpacs,
 aps,
]{revtex4-2}

\usepackage{graphicx}% Include figure files
\usepackage{dcolumn}% Align table columns on decimal point
\usepackage{bm}% bold math
\usepackage{float}
\usepackage[caption=false]{subfig}
\usepackage{subfig}
\usepackage{tabularx}
\usepackage{multirow}  % needed for some tables

\begin{document}

%\preprint{APS/123-QED}

\title{Observable proton decay and gauge coupling unification in the improved missing
doublet SU(5) model}% Force line breaks with \\
%\thanks{A footnote to the article title}%

\author{Maria Mehmood}
 \email{mehmood.maria786@gmail.com}

\author{Mansoor Ur Rehman}
 \email{mansoor@qau.edu.pk}
\affiliation{Department of Physics, Quaid-i-Azam University, Islamabad 45320, Pakistan
}

\date{\today}% It is always \today, today,
             %  but any date may be explicitly specified

\begin{abstract}
We investigate the possibility of observable proton decay within the improved missing doublet model (IMDM75), which utilizes a Higgs in 75 representation for $SU(5)$ gauge symmetry breaking.
The realization of observable proton decay in IMDM75 is made feasible primarily due to chirality nonflipping color-triplet mediation, while chirality flipping mediation is adequately suppressed. 
Our predictions suggest a range of proton lifetimes between $10^{34}-10^{36}$ yr, which can be observed in upcoming experiments such as Hyper-K and DUNE, for a corresponding range of color-triplet mass parameter $M_T$ in the order of $10^{11}-10^{14}$~GeV.  
IMDM75 is shown to offer unique predictions for branching ratios when compared to other grand unified theory models.
Finally, a realization of successful gauge coupling unification is achieved, even with the presence of particles at intermediate scales.
\end{abstract}

\maketitle

%\tableofcontents
\section{Introduction}
Proton decay  serves as a very important discriminator in distinguishing among the various models of grand unified theories (GUTs) such as $SU(5)$, flipped $SU(5)$, Pati-Salam model, and $SO(10)$, etc. Even different versions of the same model can lead to distinct predictions. 
Currently, Super-K sets a lower limit on proton lifetime \cite{Super-Kamiokande:2016exg, Super-Kamiokande:2013rwg, Super-Kamiokande:2012zik, Super-Kamiokande:2014otb, Super-Kamiokande:2005lev, ParticleDataGroup:2018ovx}, while future experiments such as JUNO \cite{JUNO:2015zny}, DUNE \cite{DUNE:2015lol, DUNE:2020ypp}, and Hyper-K \cite{Hyper-Kamiokande:2018ofw} with enhanced sensitivities are expected to distinguish between different GUT models.

The minimal supersymmetric $SU(5)$ model has been extensively studied in relation to proton decay, particularly with regard to the mediation of color triplets through dimension-five operators \cite{Hisano:1992jj, Hisano:2013exa, Nagata:2013sba, Ellis:2019fwf, Babu:2020ncc}.
However, a major challenge is the doublet-triplet splitting problem, where both electroweak doublets and color triplets in 5-plets acquire masses of similar orders. 
A successful realization of electroweak symmetry breaking requires the electroweak doublets to be sufficiently light. On the other hand, to prevent rapid proton decay, it is necessary for the color triplets to have a sufficiently heavy mass.
To address this issue, a missing partner mechanism was proposed, which involves replacing the standard GUT Higgs with a $75$-plet and adding a pair of GUT mass $50$-plets to the minimal matter content of the $SU(5)$ model \cite{Masiero:1982fe}. 
This allows the color triplets in 5-plets to mix with those in 50-plets and acquire heavy masses, while the electroweak doublets remain light. 
However, the presence of large representations like the $75$ and $50$-plets makes the model nonperturbative just above the GUT scale.
A modified version \cite{Hisano:1994fn,Hisano:1997nu} overcomes this issue by assigning Planck-scale masses to the $50$-plets and introducing additional pairs of 5 and 50-plets. 
To delve into the modified version and its variants that have undergone a phenomenologically acceptable treatment, one can refer to the following sources: 
\cite{Kounnas:1983gm, Berezhiani:1996nu,Altarelli:2000fu,Antusch:2014poa, Bajc:2016qcc,Pokorski:2019ete,Ellis:2021fhb}.

This paper explores the improved missing doublet model (IMDM75) \cite{Berezhiani:1996nu} in the context of searching for observable proton decay. 
Similar to the modified version, this improved model includes additional pairs of 5 and 50-plets.
In addition, it employs an anomalous $U(1)_A$ symmetry that solves the doublet-triplet problem without requiring any additional discrete or global symmetries. $SU(5)$ along with anomalous $U(1)_A$ symmetry can be embedded in a realistic string theory, as the existence of such an anomalous $U(1)_A$ symmetry does not imply an anomaly in the string theory \cite{Nilles:1999yv}.
The findings of this investigation show that IMDM75 predicts observable proton decay via chirality nonflipping color-triplet mediation, making it an important target for next-generation proton decay experiments.
A detailed investigation of observable proton decay in IMDM75 along with a consistent two-loop gauge coupling unification is the prime focus of this paper.

The paper is structured as follows: Section \ref{Mod} outlines the matter content and superpotential of IMDM75, along with the charge assignments under the anomalous $U(1)_A$ symmetry.
In Sec. \ref{pd}, we discuss the different modes of proton decay mediated by chirality flipping and nonflipping color triplets.
Section \ref{vsvs} provides a brief comparison of IMDM75's predictions for proton decay with those of other GUT models.
Section \ref{gcusec} describes a successful realization of gauge coupling unification, along with a discussion of the matching conditions in the presence of additional intermediate-scale particles beyond the minimal content of the minimal supersymmetric standard model (MSSM).
Our conclusions are summarized in Sec. \ref{con}.

\section{IMDM75}\label{Mod}
The matter content of MSSM and the right-handed-neutrino  superfield reside in $10_i+\bar{5}_i$ and $1_i$ representations of $SU(5)$ respectively, as shown in Table \ref{sf}. The GUT Higgs superfield is denoted by $75_H$ representation whereas the electroweak doublets arise from the mixing of $5$ and $50$-plets.
An anomalous $U(1)_A$ symmetry, broken by the nonzero vacuum expectation value (VEV) $\langle X\rangle/m_P \sim 10^{-1}-10^{-2}$ of a gauge singlet field $X$ via anomalous D term\cite{Dine:1987xk, Atick:1987gy, Dine:1987gj}, is imposed on the model, as described in \cite{Berezhiani:1996nu}.
The charge assignment under $U(1)_A$ symmetry  and $\mathbb{Z}_2$ matter parity for all superfields are provided in Table \ref{qs}. 
Note that the $X$ field has a charge of $-q$ under $U(1)_A$ and $n$ takes values $0,1,2,\cdots$. 
The $\mathbb{Z}_2$ matter parity is utilized to eliminate undesired terms in the superpotential and to obtain a potential cold dark matter candidate. 
Interestingly, the appearance of $\mathbb{Z}_2$ matter parity can be an accidental by-product of the $U(1)_A$ symmetry breaking, as discussed in \cite{Berezhiani:1996nu}. 
It is noteworthy that for $q=1$ and $n=2$, only the $\mathbb{Z}_7$ subgroup of the $U(1)_A$ symmetry is sufficient to restrict the model. This implies a complete breakdown of the $U(1)_A$ symmetry once $X$ obtains a VEV, eliminating any potential domain wall problems. An additional set of $5$ and $50$-plets ($5',\ \bar{5}', 50',\ \overline{50}'$) is necessary to prevent rapid proton decay while maintaining gauge coupling unification (GCU), as explained in \cite{Berezhiani:1996nu}.
\begin{table}[t!]
\begin{center}
\caption{\label{sf} The superfield content of $SU(5)$ and corresponding decomposition under the MSSM gauge group.}
\begin{small}
\begin{tabular}{|>{\centering\arraybackslash}m{1.3cm}|>{\centering\arraybackslash}m{6.8cm}| }
\hline
%\vspace{0.1cm}
$SU(5)$&$SU(3)_c\times SU(2)_L \times U(1)_Y$\\
\hline   
%\vspace{0.05cm}
 $10_i$  & $Q_i ( 3, 2, 1/6)$
 + $U^c_i ( \overline{3}, 1, -2/3)$  
 + $E^c_i ( 1, 1, 1)$ \\
$ \overline{5}_i$ & $D^c_i ( \overline{3}, 1, 1/3)$
+ $L_i ( 1, 2, -1/2)$\\
$ 1_i$ & $N^c_i ( 1, 1, 0)$  \\
\hline
%\vspace{0.2cm}
 $5_{h}$  & $H_{u}(1,2,+1/2)$ 
 + $H_{T}(3,1,-1/3)$  \\
 $\bar{5}_{h}$  & $ {H}_{d}(1,2,-1/2)$  
 + $\bar{H}_{T}(\bar{3},1,+1/3)$  \\
 $75_H$  & $(\bar{3},1,-5/3)$ 
 + $ (3,1,5/3)$ 
 + $ (3,2,-5/6)$ \\
 &+ $ (\bar{3},2,5/6)$ 
 + $ (6,2,5/6)$  
 + $ (\bar{6},2,-5/6)$ \\
 &+ $ S(1,1,0)$ 
 + $ (8,1,0)$ 
 + $ (8,3,0)$\ \ \ \ \ \ \ \ \ \ \ \\
 $24_A$  & $(1,1,0)$ 
 + $(1,3,0)$ 
 + $(8,1,0)$ \\
 &+ $\chi(3,2,-5/6)$ 
 + $\bar{\chi}(\bar{3},2,5/6)$ \\
 $ 50$  & $(3,1,-1/3)$ 
 + $ (\bar{3},2,-7/6)$ 
 + $ (1,1,-2)$ \\
 &+ $ (6,1,4/3)$ 
 + $ (\bar{6},3,-1/3)$ 
 + $ (8,2,1/3)$ \\
 $\overline{50}$  & $(\bar{3},1,+1/3)$ 
 + $ (3,2,+7/6)$ 
 + $ (1,1,+2)$ \ \ \ \ \ \  \\
 & + $ (\bar{6},1,-4/3)$ 
 + $ (6,3,+1/3)$ 
 + $ (8,2,-1/2)$ \\
\hline
\end{tabular}
\end{small}
\end{center}
\end{table}

The superpotential for the IMDM75 model with $U(1)_A \times \mathbb{Z}_2$ symmetry, as shown in Table \ref{qs}, is given by 
\begin{eqnarray} \label{superpot}
W&=&\frac{M}{2}\, 75^2_H -\frac{\lambda_{75}}{3}\, 75^3_H \nonumber \\
&+&\frac{1}{8}y^{(u)}10\, 10\, 5_h  -  y^{(l,d)}  10\, \bar{5}\, \bar{5}_h \nonumber\\
&+& {\lambda}\, \overline{50}\, 75_H\, 5_h+{\lambda'}\, \overline{50}'\, 75_H\, 5'\nonumber \\
& +&{\bar{\lambda} \, \bar{5}_h\, 75_H\ 50'}+{\bar{\lambda}' \, \bar{5}'\, 75_H\ 50}\nonumber\\
&+&\lambda_{50}\, \langle X\rangle\, \overline{50}\, 50 +\lambda'_{50}\, \langle X\rangle\, \overline{50}'\, 50' \nonumber\\
&+& y^{(\nu)}\bar{5}\,1\, 5_h + M_R 1\, 1 + W_{NR},
\end{eqnarray}
where generation indices have been suppressed. 
The first two terms in $W$ break the $SU(5)$ symmetry and attain the VEV of $\langle 75_H \rangle \equiv \upsilon = 3M/4\lambda{75}$ in the singlet direction of MSSM gauge symmetry  $SU(3)_c\times SU(2)_L\times U(1)Y$. 
The Yukawa interactions of charged superfields are described by the terms in the second line, while the third and fourth lines provide heavy masses to the color triplets in 5- and 50-plets and leaving the electroweak doublets in 5-plets massless. 
This is the missing partner mechanism, which solves the doublet-triplet splitting problem without requiring parameter fine-tuning.
The fifth line provides heavy masses of the order of $\sim 10^{17}-10^{18}$~GeV  to 50-plets while keeping the model perturbative above the GUT scale. 
The first two terms in the last line explain tiny neutrino masses through the seesaw mechanism \cite{Mohapatra:1986bd}. The last term, $W_{NR}$, includes all relevant nonrenormalizable terms that will be discussed later.
\begin{table}[t!]
\begin{center}
\caption{\label{qs} Superfields with $\mathbb{Z}_2$ and $U(1)_A$ charges.}
\begin{tabular}{|>{\centering\arraybackslash}m{0.8cm}|>{\centering\arraybackslash}m{0.8cm}|>{\centering\arraybackslash}m{2.0cm}| }
\hline
$SU(5)$&$Q_{\mathbb{Z}_2}$&$5\times Q_{ A}$ \\
\hline 
%\vspace{0.1cm}
 $10_i$  & $1 $ &$ (n+3)q$   \\
$ \overline{5}_i$  & $1$ &$2(n+3)q$ \\
$ 1_i $ & $1$& $0$ \\
 $5_h $ &$0$&$-2(n+3)q$  \\
 $\overline{5}_h$ &$0$& $-3(n+3)q$   \\
  $5' $ &$0$&$(3n+4)q$  \\
 $\overline{5}'$ &$0$& $(2n+1)q$ \\
\hline
\end{tabular}
\begin{tabular}{|>{\centering\arraybackslash}m{0.8cm}|>{\centering\arraybackslash}m{0.8cm}|>{\centering\arraybackslash}m{2.0cm}| }
\hline
$SU(5)$&$Q_{\mathbb{Z}_2}$&$5\times Q_{ A}$\\
\hline  
%\vspace{0.1cm}
$ 75_H$ & $0$&$0$ \\
$X$&$0$&$-5q$\\ 
$\quad$ & $\quad$ & $\quad$ \\ 
$50$&$0$&$-(2n+1)q$\\ 
$\overline{50}$&$0$&$2(n+3)q$\\ 
$50'$&$0$&$3(n+3)q$\\ 
$\overline{50}'$&$0$&$-(3n+4)q$\\
\hline
\end{tabular}
\end{center}
\end{table}

The $U(1)_A$ symmetry in the IMDM75 model forbids the $\mu\ 5_h\bar{5}_h$ term, which leads to the electroweak doublets in $5_h$ and $\bar{5}_h$ remaining massless up to all orders.
To generate the MSSM $\mu$ term for the light electroweak doublets, the Giudice-Masiero mechanism \cite{Giudice:1988yz} can be utilized, which is a consequence of supergravity breaking in the hidden sector. 
The term $y^{(l,d)}_{ij}10_i\ \bar{5}_j\bar{5}_h$ leads to same mass matrix for down type quark and charges leptons. This can be avoided by using nonrenormalizable term
\begin{eqnarray}
W_{NR} &\supset & \frac{\kappa_{ij}}{m_P}75_H\ 10_i\ \bar{5}_j\ \bar{5}_h
\end{eqnarray}
Yukawa terms for quarks and charged leptons in superpotential of IMDM75 include
\begin{eqnarray}
W&\supset &\frac{1}{8}y^{(u)}_{ij}\ 10_i\ 10_j\ 5_h -y^{(l,d)}_{ij}\ 10_i\ \bar{5}_j\ \bar{5}_h \nonumber \\
&+& \frac{\kappa_{ij}}{\Lambda}75_H\ 10_i\ \bar{5}_j\ \bar{5}_h
\end{eqnarray}
where $i$ and $j$ are generation indices. After diagonalizing Yukawa couplings and calculating mass eigenstates, down-type quark and charged lepton Yukawa couplings are redefined as
\begin{eqnarray}
Y^{(l)}_{ij}&=& y^{(l,d)}_{D\ ij}-\frac{3\upsilon \kappa_{ij}}{m_P}V^T \\
Y^{(d)}_{ij}&=& y^{(l,d)}_{D\ ij}
+\frac{\upsilon \kappa_{ij}}{m_P} V^T 
\end{eqnarray}
The Yukawa terms in the effective superpotential can be expressed in terms of mass eigenstates as
\begin{eqnarray}
W_{eff}&\supset &
Y^{(u)}_{ij} U^c_{b\ i} (Q^{b}_j. H_u)
-V^* Y^{(d)}_{ij}(Q^{a}_i.H_d)D^c_{a\ j}\nonumber \\
& -&Y^{(l)}_{ij}E^c_i ( L_j . {H}_d)-\frac{1}{2}\epsilon_{a b c}e^{i\varphi_i} Y^{(u)}_{ij}  (Q^{a}_i . Q^{b}_j)\  H^{ c}_T \nonumber \\
&+&(Y^{(u)}_{ij} V) U^c_{a\ i}\ E^c_j\ H^{a}_T
+V^*Y^{(d)}_{ij} (Q^{a}_i . L_j )\bar{H}_{T a} \nonumber \\
&+&\epsilon^{abc} e^{-i\varphi_i} V^*Y^{(d)}_{ij}D^c_{a\ i}U^c_{b\ j}\bar{H}_{T\ c}
\end{eqnarray}   
here $a,b,c,...$ are $SU(3)_c$ color indices. $Y^{(u)}, Y^{(d)}, Y^{(l)}$ are diagonalized Yukawa couplings effectively.
Furthermore, fermion mass hierarchy is explained by introducing a generation-dependent horizontal $U(1)_A$ symmetry and incorporating nonrenormalizable terms from $W_{NR}$  in \cite{Berezhiani:1996nu}.
\section{Proton decay in IMDM75\label{pd}}
Proton decay can occur via the color triplets through dimension-five and -six operators. 
On the other hand, GUT gauge bosons can mediate proton decay only through dimension-six operators at the leading order. Chirality flipping and nonflipping operators contribute to both dimension-five and -six levels. However, we can observe testable predictions only through the chirality nonflipping operators in color-triplet mediation as explained below.
\subsection{Dimension-five proton decay}
\begin{figure}[t!]\centering
\subfloat[\label{LLLL}$Q\ Q\ Q\ L$]{\includegraphics[width=1.20in]{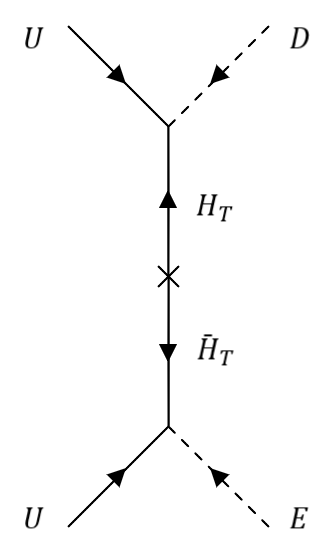}} \quad
\subfloat[\label{RRRR}$U^c D^c U^c E^c$]{\includegraphics[width=1.24in]{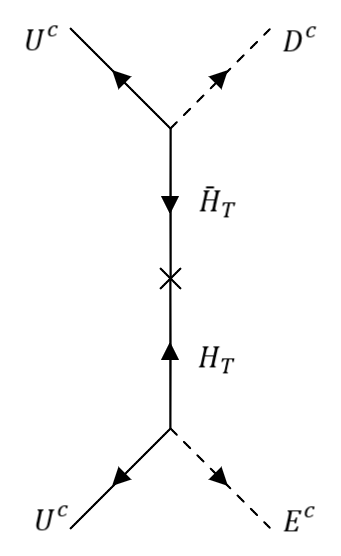}}
\caption{\label{dim5}Dimension-five two fermion two scalar proton decay operators mediated via chirality flipping color triplets of $5_h$ and $\bar{5}_h$.}
\end{figure}

A detailed analysis of proton decay originated from the dimension-five operators is in order. In IMDM75 the renormalizable operators involving color triplets $(H_T,\bar{H}_T) \subset (5_h, \bar{5}_h)$ are
\begin{eqnarray}
y^{(u)} 10\, 10\, 5_h& \supset & y^{(u)} U D H_T+y^{(u)} U^c E^c H_T,\\
y^{(l,d)} 10\, \bar{5}\, \bar{5}_h& \supset & y^{(d)} D N \bar{H}_T + y^{(d)} U^c D^c \bar{H}_T\nonumber \\
&+& y^{(l)} U E \bar{H}_T\, .
\end{eqnarray}
These Yukawa interactions along with chirality flipping mass term, $5_h\bar{5}_h$, lead to dimension-five proton decay operator with two scalars and two fermions, as shown in Fig.~\ref{dim5}.
If allowed the chirality flipping mass term, $5_h\bar{5}_h$,  leads to fast proton decay even with color-triplet masses of the order of GUT scale, thanks to the $U(1)_A$ symmetry which forbids this term in IMDM75 up to all orders. This term, however, can arise indirectly via $U(1)_A$ symmetry-breaking interactions induced by the VEV $\langle X\rangle$, as explained below. 

\begin{figure}[t!]\centering
\subfloat[]{\includegraphics[width=2.0in]{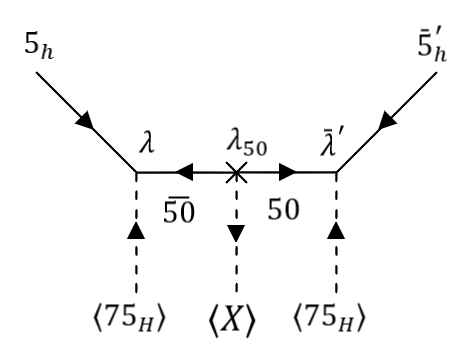}}\quad
\subfloat[]{\includegraphics[width=1.5in]{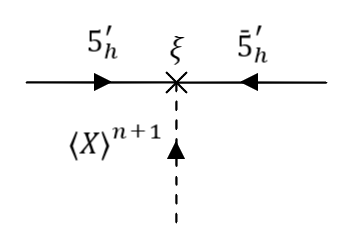}}\quad
\subfloat[]{\includegraphics[width=2.0in]{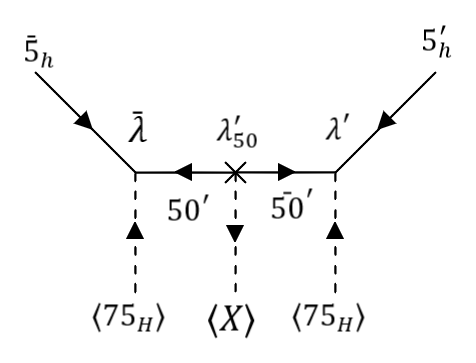}}
\caption{\label{ct-mass}Interactions of color triplets in primed and unprime 5- and 50-plets.}
\end{figure}

\begin{figure}[t!]\centering
{\includegraphics[width=0.45\textwidth]{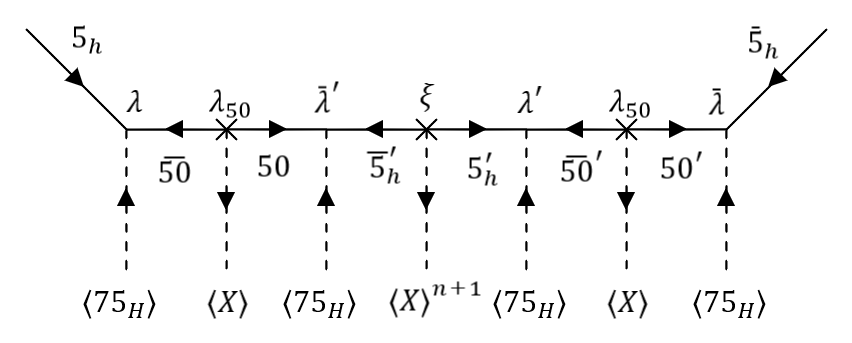}}
\caption{\label{5h5hb} Feynman diagram representing interactions leading to the effective $5_h\bar{5}_h$ term for color triplets.}
\end{figure}

As a matter of convenience, the mixing terms of the color triplets among, $5_h,\bar{5}_h,5',\bar{5}',\,50,\,\overline{50},\,50'$ and $\overline{50}'$ in Eq.~(\ref{superpot}) are represented diagrammatically in Fig.~\ref{ct-mass}. 
Especially, the effective mass term, $5'\bar{5}'$, is generated via $\langle X\rangle$ as
\begin{eqnarray}
W_{NR} \supset  M'_D\, 5'\, \bar{5}',
\end{eqnarray}
where
\begin{eqnarray}
M'_D \equiv   \frac{\xi \langle X\rangle^{n+1}}{m_P^n},
\end{eqnarray}
and $m_P \simeq 2.4 \times 10^{18}$~GeV is the reduced Planck mass. 
This term ensures that the electroweak doublets and the color triplets in $5'$-plets acquire the same masses, disregarding any mixing terms of the color triplets. Moreover, it results in an effective term, $5_h\bar{5}_h$, as depicted in Fig.~\ref{5h5hb}.
However, it is important to note that the effective term, $5_h\bar{5}_h$, couples only the color triplets in $5_h\ \bar{5}_h$ and not the electroweak doublets, as there are no electroweak doublets in 50-plets. 

The mass matrix relevant for chirality flipping color-triplet mediation is given by
\begin{eqnarray}\label{mttbar}
\mathcal{M}_{T\overline{T}}=
\begin{pmatrix}
0  & 0 & \lambda \ \upsilon & 0\\
0 & {\xi \langle X\rangle^{n+1}}/{m_P^n} & 0 &{\lambda}' \ \upsilon \\
0 & \bar{\lambda}' \ \upsilon &\lambda_{50} \langle X\rangle & 0\\
\bar{\lambda} \ \upsilon & 0 & 0 & \lambda_{50}' \langle X\rangle
\end{pmatrix},
\end{eqnarray}
where the mass matrix element relevant for proton decay via chirality flipping term $5_h\bar{5}_h$ is
\begin{eqnarray}\label{meff}
\frac{1}{(\mathcal{M}_{T\overline{T}})^{-1}_{\ 11}}
 & = &  \lambda \bar{\lambda} \lambda' \bar{\lambda}' \, 
\frac{\upsilon^4}{M'_D M_{50}^{2}}.
\end{eqnarray}
The dimension-five proton decay interactions contribute in a loop diagram, which generates the effective dimension-six four-Fermi proton decay operator.
The decay rate, aside from loop factors, varies as
\begin{eqnarray}
\Gamma
&\propto &  \frac{\mu \ [(\mathcal{M}_{T\overline{T}})^{-1}_{ \ 11}]^2}{M'_D},
\end{eqnarray}
where the mass $\mu$ arises from the $\mu$ term of the GM mechanism. We obtain $1/[(\mathcal{M}_{T\overline{T}})^{-1}_{\  11}]\sim 10^{16}$ GeV and $\mu/M'_D\sim 10^{-10}$ for typical parameter values. Thus, we conclude that proton decay in IMDM75 through chirality flipping dimension-five operators is safe from rapid proton decay.

\begin{figure}[t!]\centering
\includegraphics[width=0.40\textwidth]{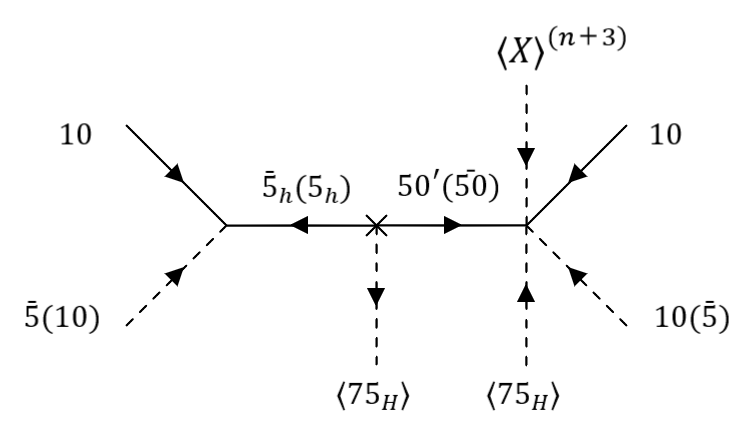}
\caption{\label{dim5hetro} Feynman diagram representing the heterogeneous proton decay operator.}
\end{figure}

Next, we examine a different type of dimension-five operators for proton decay, which arise due to the intermixing of renormalizable and nonrenormalizable interactions from $W_{NR}$ as follows
\begin{eqnarray}
W_{NR} \supset  
\frac{75_H}{m_P}\frac{X^{(n+3)}}{m_P^{(n+3)}}\left(\eta_1 50' 10\, 10
+ \eta_2 \overline{50}\, 10\, \bar{5}\right).
\end{eqnarray}
We will refer to the dimension-five proton decay operators, which result from the mixing of renormalizable and nonrenormalizable interactions in $W_{NR}$, as heterogeneous operators. The reason for this is that the color triplets involved in these operators originate from different representations of $SU(5)$, namely, the $5$- and $50$-plets. Figure \ref{dim5hetro} depicts these operators diagrammatically.
Although potentially dangerous for proton decay, these heterogeneous operators do not significantly contribute to the decay rate due to an extra suppression factor. Specifically, for $n \geq 1$, this factor is $(\langle X\rangle /m_P)^{(n+3)}(\upsilon /m_P) \lesssim 10^{-7}$. It is worth noting that the contribution of the dimension-five proton decay operator, which is induced by a Planck-scale suppressed operator $10_i 10_j 10_k \bar{5}_l$, is suppressed by the $U(1)_A$ symmetry, as also discussed in \cite{Berezhiani:1996nu}.
\subsection{Observable chirality nonflipping dimension-six proton decay}
\begin{figure}[t!]\centering
\subfloat[$(Q\ U^{c\dagger})\ (Q\ E^{c\dagger})$]{\includegraphics[width=1.40in]{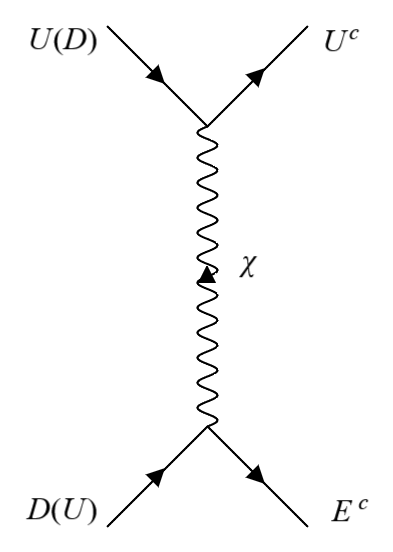}} \quad
\subfloat[$(Q\ U^{c\dagger})\ (L\ D^{c\dagger})$]{\includegraphics[width=1.4in]{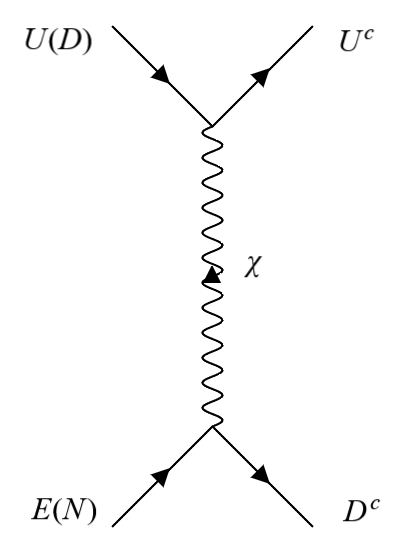}}
\caption{\label{gb}$LLRR$ type dimension-six (four-fermion) proton decay operators mediated via gauge boson $24_A\supset \chi$.}
\end{figure}

Proton decay induced by the dimension-six operators is mediated through both the gauge bosons  $(\chi, \bar{\chi}) \subset 24_A$ and the color triplets $(H_T,\bar{H}_T) \subset (5_h, \bar{5}_h)$. 
%============= gauge boson mediation =========
The interaction terms for the gauge boson mediation arise from the following part of the K\"ahler potential
\begin{eqnarray}
K&\supset &
\sqrt{2}g_5\left[-(D^{c\dagger})^{a}\bar{\chi}_a^m L_m
+\epsilon^{abc}Q^{\dagger}_{a m}\bar{\chi}^m_b P^{\dagger}U^c_c \right. \nonumber \\
&+&\left.\epsilon_{mn}E^{c\ \dagger}V^{\dagger}\bar{\chi}^m_a Q^{a n}+h.c\right],
\end{eqnarray}
where $P$ is a diagonal matrix such that $det(P)=1$ and $V$ represents the Cabibbo-Kobayashi-Maskawa (CKM) matrix. Feynman diagrams for gauge boson mediated proton decay are shown in Fig.~\ref{gb}.
These interactions correspond to proton decay operators of the $LLRR$ chirality type.

%=========== color triplet mediation ==============

\begin{figure}[t!]\centering
\subfloat[\label{2f2s1}$Q\ Q\ (U^cE^c)^{\dagger}$]{\includegraphics[width=1.30in]{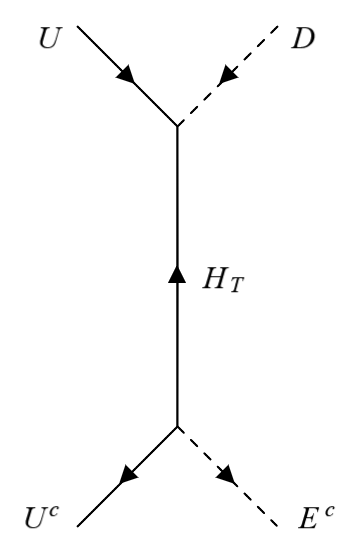}}\quad
\subfloat[\label{2f2s2}$Q\ L\ (U^c D^c)^{\dagger}$]{\includegraphics[width=1.5in]{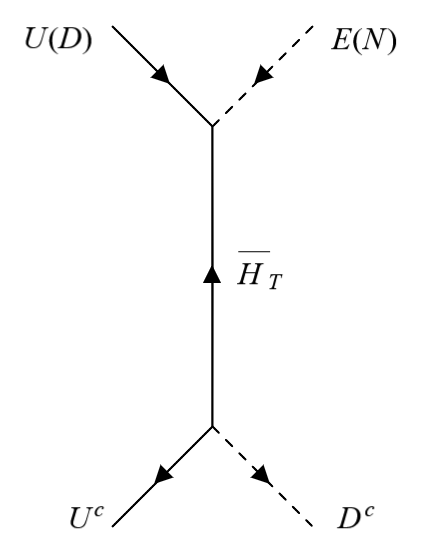}}\\
\subfloat[\label{4f1}$Q\ Q\ (U^cE^c)^{\dagger}$]{\includegraphics[width=1.25in]{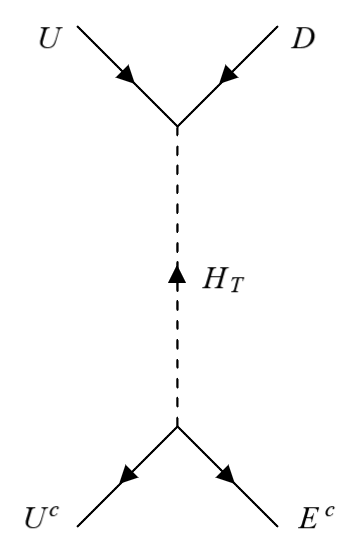}}\quad
\subfloat[\label{4f2}$Q\ L\ (U^c D^c)^{\dagger}$]{\includegraphics[width=1.45in]{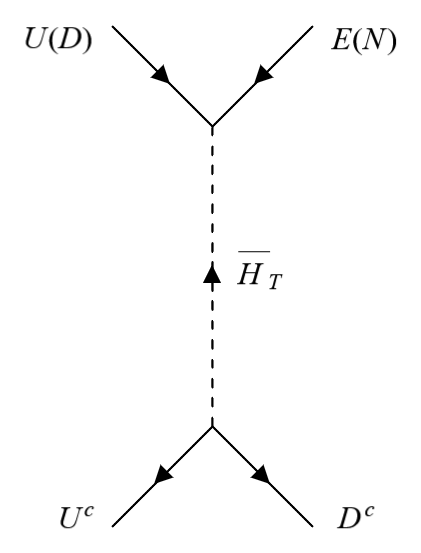}}
\caption{\label{dim6ct}$LLRR$ type dimension-six proton decay operators mediated via color triplets of $5_h,\ \bar{5}_h$.}
\end{figure}

The proton decay operators of $ LLRR $ chirality type, mediated by color triplets, can be depicted through the diagrams in Fig.~\ref{dim6ct} for dimension-six.
Even though Figs.~\ref{2f2s1} and \ref{2f2s2} have two fermions and two scalars, they originate from the dimension-six operators, as clarified in \cite{Mehmood:2020irm}.
These operators undergo further dressing in the box diagram to generate the four-Fermi proton decay operators effectively. 
Hence, their contribution to the proton decay rate is expected to be comparatively suppressed by loop factors as compared to tree-level four-fermion diagrams shown in Figs.~\ref{4f1} and \ref{4f2}. 
Consequently, in the subsequent discussion, only the four-fermion diagrams of Figs.~\ref{4f1} and \ref{4f2} will be considered for chirality nonflipping color-triplet mediation of type $LLRR$.

For chirality nonflipping dimension-six operators we need the corresponding interaction mass matrix for the color triplets.
%================== 5hbar mass============
\begin{figure}[t!]\centering
{\includegraphics[width=0.45\textwidth]{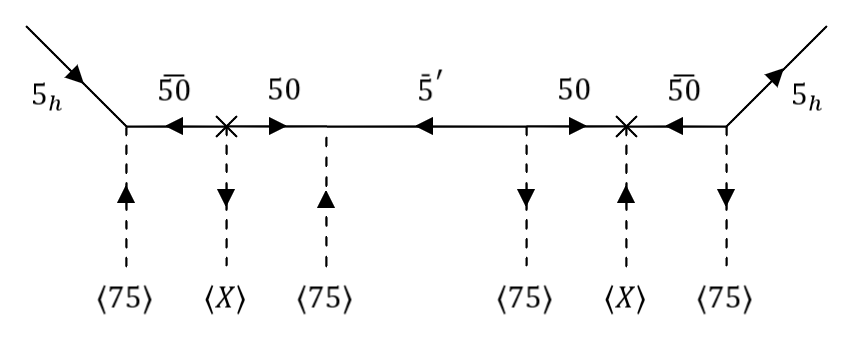}}
\caption{\label{5h} Feynman diagram representing interactions leading to interaction mass of a color-triplet in $5_h$. A similar diagram can be drawn for a color-triplet in $\bar{5}_h$}
\end{figure}
The mass matrix relevant for chirality nonflipping color-triplet mediation is
\begin{eqnarray}
\mathcal{M}_{T}&=&
\begin{pmatrix}
0  & 0 & 0 & \lambda \ \upsilon \\
0 & 0 & \bar{\lambda}' \ \upsilon & 0  \\
0 & \bar{\lambda}' \ \upsilon & 0 &\lambda_{50} \langle X\rangle \\
{\lambda} \ \upsilon  & 0 & \lambda_{50} \langle X\rangle & 0
\end{pmatrix}.
\end{eqnarray}
Figure \ref{5h} illustrates a diagram for fermionic color-triplet mediation involving $5_h$. The relevant mass matrix element for the mediation of $H_T\subset 5_h$ can be expressed as follows
\begin{eqnarray}
\sqrt{\frac{1}{(\mathcal{M}^{2}_{T^*T})^{-1}_{\  11}}} 
&=& \sqrt{\frac{|\lambda|^2 |\bar{\lambda}'|^2  \upsilon^4 }{M_{50}^2 + |\bar{\lambda}'|^2 \upsilon^2}}\, , \nonumber \\
&\simeq & \frac{\lambda \bar{\lambda}' \upsilon^2}{M_{50}}\, . \label{HT}
\end{eqnarray}
Here, it is assumed that the couplings $\lambda$ and $\bar{\lambda}'$ are real, $M_{50}=\lambda_{50}\langle X \rangle$, and
\begin{eqnarray}
\mathcal{M}^2_{T^*T}  &=& \mathcal{M}^*_{T} \mathcal{M}_{T}.
\end{eqnarray} 
%================== 5hbar mass============
Similarly for the mediation of $\bar{H}_T \subset \bar{5}_h$, we obtain
\begin{eqnarray}
\sqrt{\frac{1}{(\mathcal{M}^2_{\overline{T}^*\overline{T}})^{-1}_{ \ 11}}} 
&\simeq & \frac{\bar{\lambda}{\lambda}' \upsilon^2}{M_{50}}\, , \label{HbT}
\end{eqnarray}
where we assume couplings $\bar{\lambda},\ {\lambda}'$ to be real. We also assume $\lambda'_{50}\langle X \rangle=\lambda_{50}\langle X \rangle=M_{50}$. For $M_{50} \sim 2 \times 10^{17}\ \text{GeV}$,  $\upsilon \sim 6 \times 10^{15}\ \text{GeV}$ and $(\lambda, \bar{\lambda}',\bar{\lambda},{\lambda}' )\sim \lambda$ we get
\begin{small}
\begin{eqnarray}
\sqrt{\frac{1}{(\mathcal{M}^{2}_{T^*T})^{-1}_{\  11}}} 
\sim
\sqrt{\frac{1}{(\mathcal{M}^2_{\overline{T}^*\overline{T}})^{-1}_{ \ 11}}} 
\sim 
{\lambda}^2 \,( 1.6  \times 10^{14} \text{GeV}).
\end{eqnarray}
\end{small}This mediation leads to observable proton decay for $0.03 \lesssim \lambda \lesssim 0.2$ as shown below. 

\begin{figure*}[t]\centering
\subfloat[$p \rightarrow e^+ \pi^0$]{\includegraphics[width=0.45\textwidth]{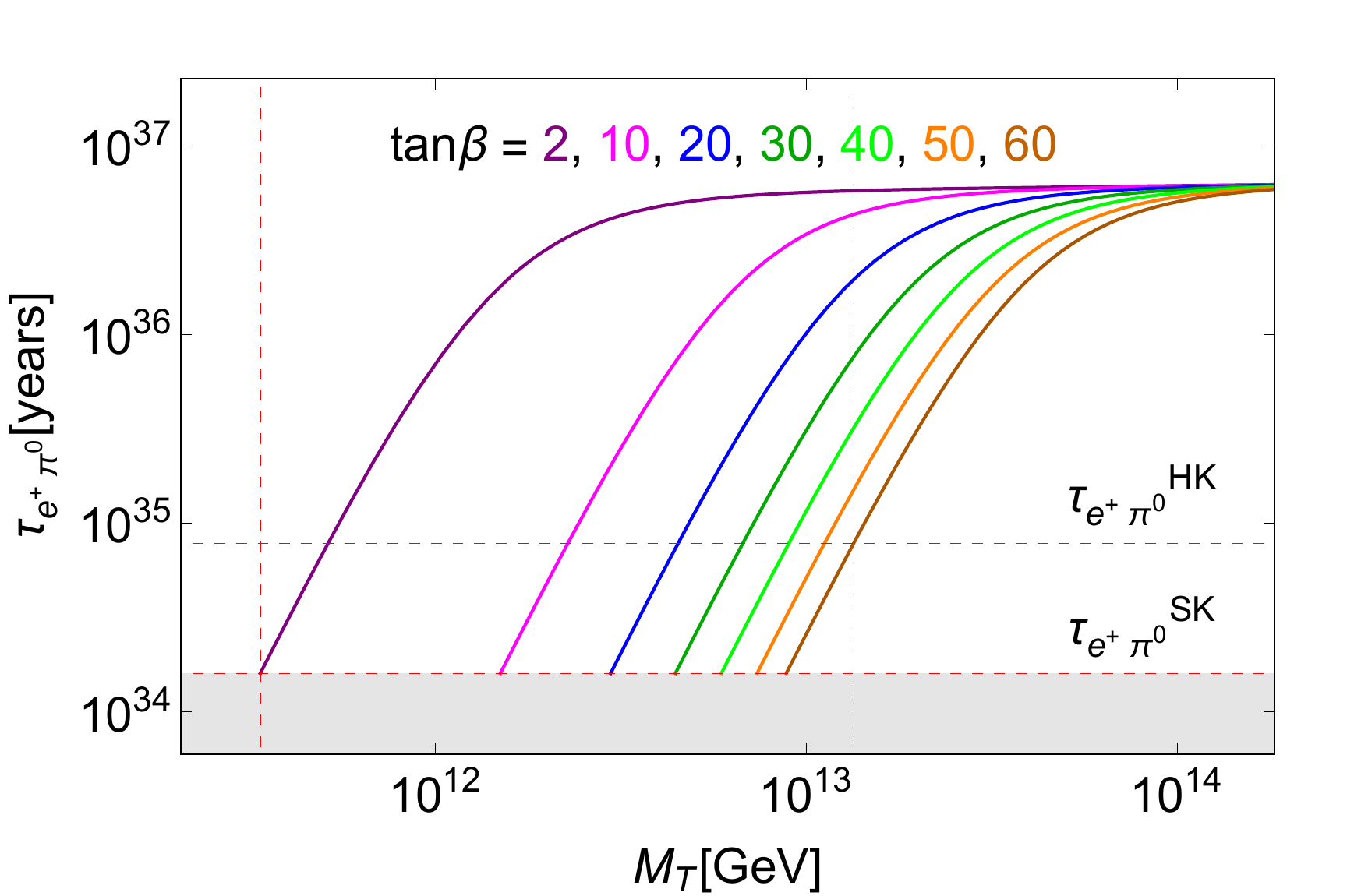}}
\subfloat[$p \rightarrow \mu^+ \pi^0$]{\includegraphics[width=0.45\textwidth]{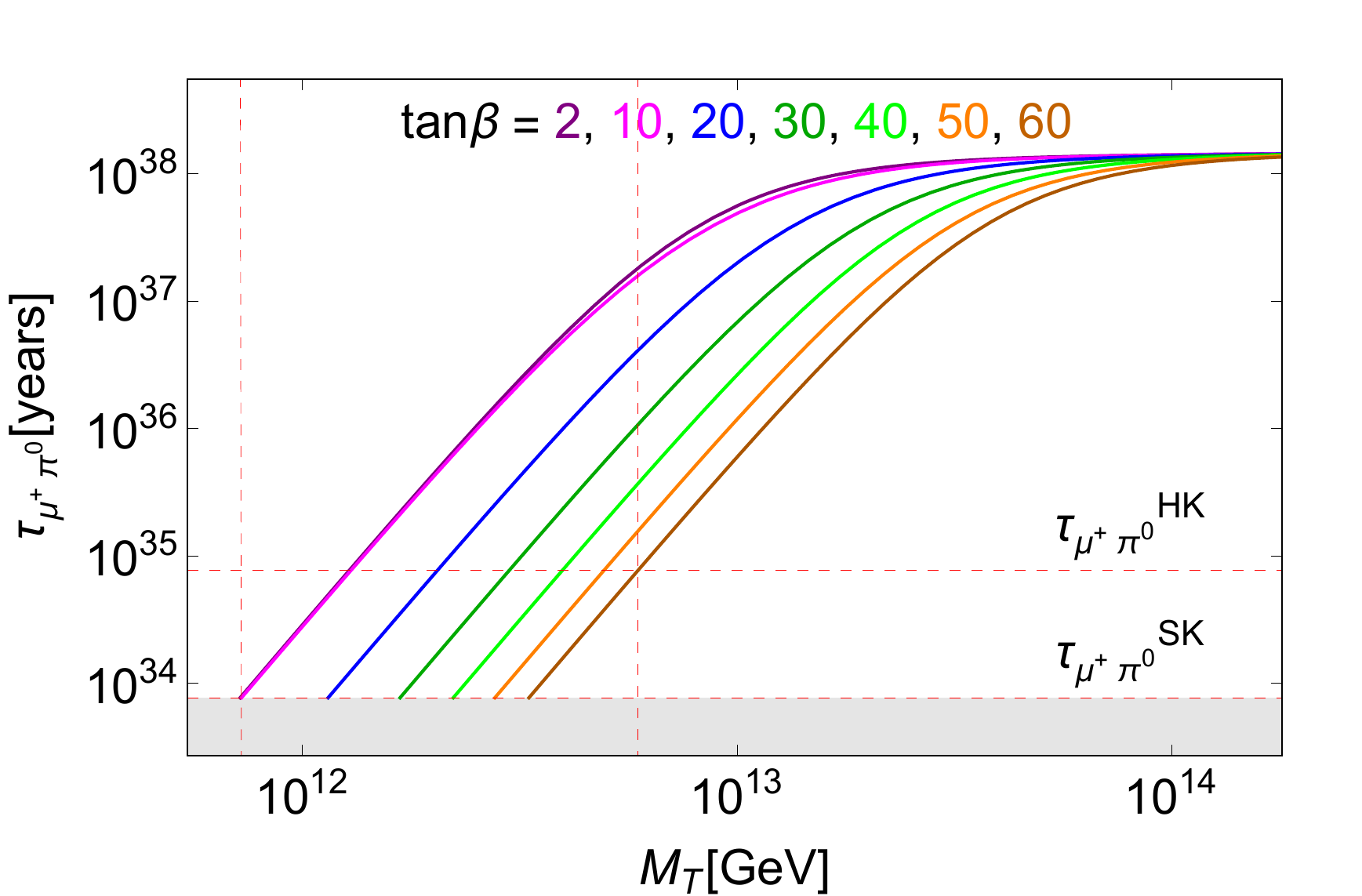}}\\
\subfloat[$p \rightarrow e^+ K^0$]{\includegraphics[width=0.45\textwidth]{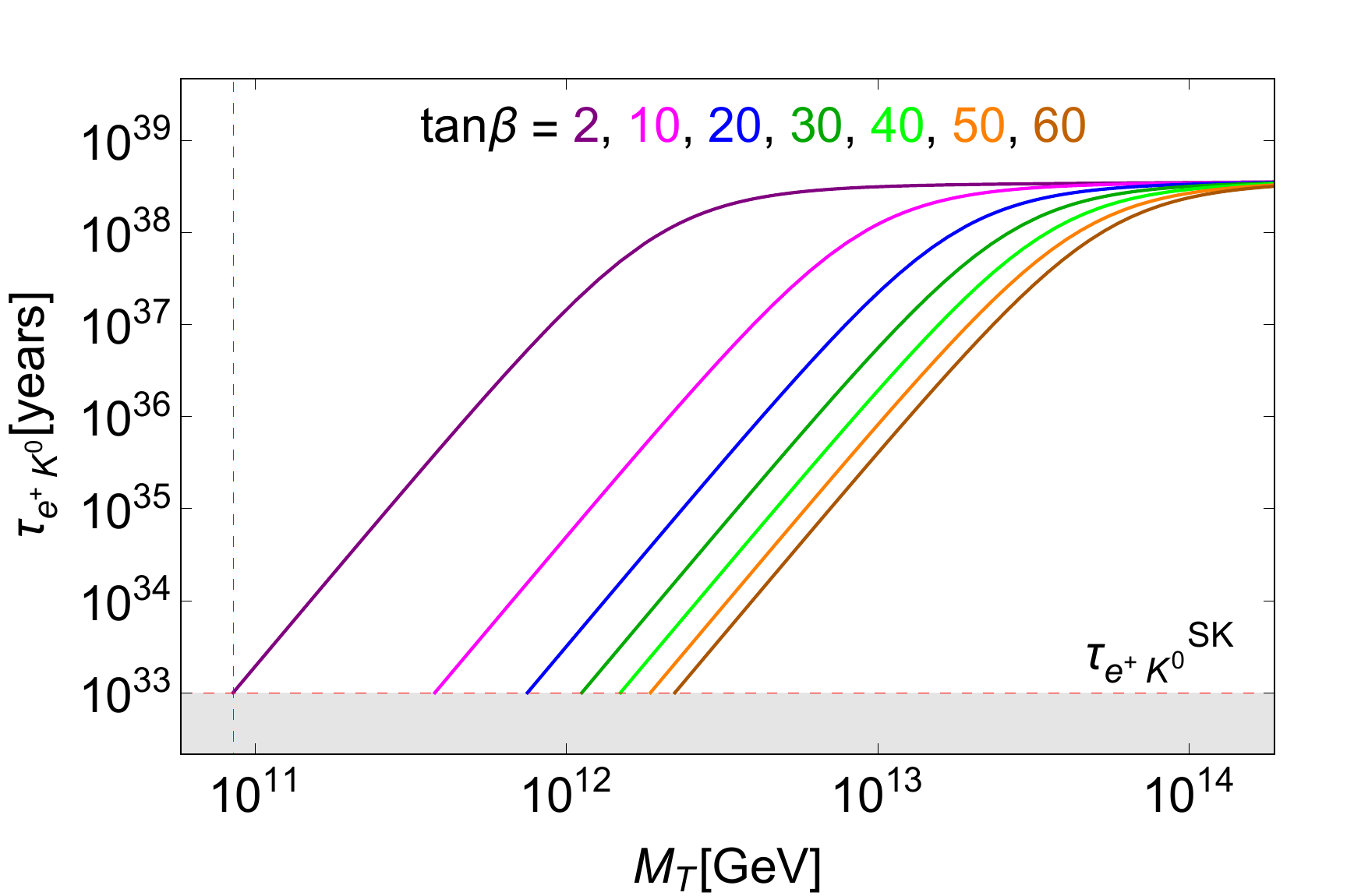}}
\subfloat[$p \rightarrow \mu^+ K^0$]{\includegraphics[width=0.45\textwidth]{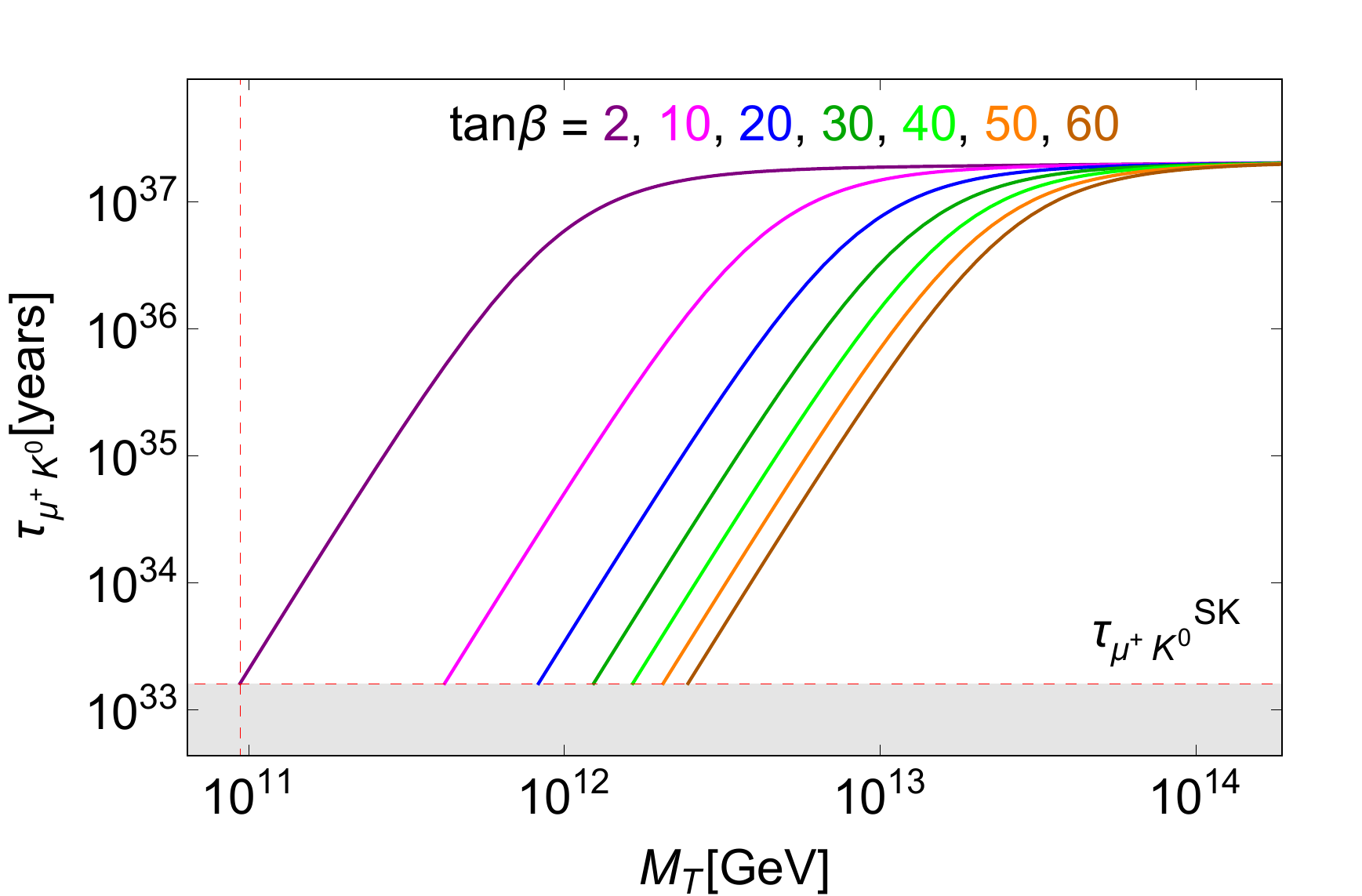}}
\caption{\label{clpd} The partial-lifetime estimates of proton for charged-lepton decay channels as a function of  $M_T $ with $\tan \beta$ in the range, $2\leq\tan{\beta}\leq 60$ (increasing from left to right). The bottom dashed lines represent the experimental limits from Super-K and the top dashed lines represent Hyper-K limits. }
\end{figure*}

%============ operators and Wilson's coefficients =====
The contributions of chirality nonflipping mediation via color triplets and gauge bosons are presented in the following dimension-six  proton decay operators
\begin{eqnarray}
\mathcal{L}_6& \supset &
C^{ijkl}_{6(1)} \mathcal{O}_{6(1)}
+C^{ijkl}_{6(2)} \mathcal{O}_{6(2)}\, ,
\end{eqnarray}
where $\mathcal{O}_{6(i)}$ are the dimension-six proton decay operators defined as

\begin{eqnarray}
\mathcal{O}_{6(1)}
&=& \int d^2\theta d^2\bar{\theta} e^{-i\varphi_k}V_{lp}(U^{c}_i)^{\dagger}(D^{c}_j)^{\dagger}U_kE_p\nonumber \\
&+& \int d^2\theta d^2\bar{\theta}(U^{c}_i)^{\dagger}(D^{c}_j)^{\dagger} D_k N_l\, ,\\
\mathcal{O}_{6(2)}
&=&2\int d^2\theta d^2\bar{\theta}  e^{-i\varphi_i}U_i D_j(U^{c}_k)^{\dagger}(E^{c}_l)^{\dagger}\, ,
\end{eqnarray}
and $C^{ijkl}_6$ are Wilson's coefficients given by

\begin{eqnarray}
C^{ijkl}_{6(1)}
&=&-e^{i\varphi_i}( Y^{(d)*} V^T )_{ij}(V^* Y^{(d,e)})_{kl}(\mathcal{M}^2_{\overline{T}^*\overline{T}})^{-1}_{\ 11} \nonumber \\
&-&\frac{g^2_5}{M^2_{\chi}}e^{i\varphi_i}\delta_{ki}\delta_{jl}\, ,\\
C^{ijkl}_{6(2)}
&=&-\frac{1}{2}e^{i\varphi_i}Y^{(u)}_{ij}\delta_{ij}
 V^{\dagger}_{kl}Y^{(u) *}_{ll}\ (\mathcal{M}^2_{T^*T})^{-1}_{\ 11}\nonumber \\
&-&\frac{g^2_5}{M^2_{\chi}}e^{i\varphi_i}\delta_{ik}V^{\dagger}_{lj}\,.
\end{eqnarray}
Here $\phi_i$ is the phase angle, $Y^{(x)}$ represents diagonalized Yukawa couplings, $V$ represents the CKM matrix, $g_5$ is $SU(5)$ gauge coupling at GUT scale, and $M_{\chi}=\sqrt{24}\, g_5\, \upsilon$ is the mass of gauge bosons.
%========= Decay rates ====================
%========= charged lepton channels ========
The decay rates for charged lepton channels are
\begin{widetext}
\begin{eqnarray}
\Gamma_{l^+ \pi^0} &=& 
\mathit{F}_{\pi^0}T^2_{l^+ \pi^0}\left(A_{S_1}^2 \left|\frac{g^2_5}{M^2_{\chi}} V_{1p} +Y^{( d)2}_{11} V_{1p} (\mathcal{M}^2_{\overline{T}^*\overline{T}})^{-1}_{\ \ 11}\right|^2 
+ A_{S_2}^2\left|\frac{g^2_5}{M^2_{\chi}}V^*_{1p}+Y^{(u)}_{11} Y^{(u)}_{pp} V^{*}_{p1}(\mathcal{M}^2_{T^*T})^{-1}_{\ 11}\right|^2  \right) , \\
\Gamma_{l^+ K^0} &=&
\mathit{F}_{K^0}T^2_{l^+ K^0} \left( A_{S_1}^2\left|\frac{g^2_5}{M^2_{\chi}}V_{2p} +Y^{( d)}_{11}Y^{( d)}_{mm}V_{21}V^*_{1m}V_{mp}\ (\mathcal{M}^2_{\overline{T}^*\overline{T}})^{-1}_{\ 11}\right|^2 
+A_{S_2}^2\left|\frac{g^2_5}{M^2_{\chi}}V^{*}_{2p}\right|^2\right) ,
\end{eqnarray}
\end{widetext}
where
\begin{eqnarray}
\mathit{F}_x &=& \frac{m_p A^2_L}{32\pi}\left(1-\frac{m^2_{x}}{m^2_p}\right)^2,
\end{eqnarray}
\begin{table*}[t]
\begin{center}
\caption{The Super-K bounds, Hyper-K, and DUNE sensitivities and values of relevant matrix elements for various proton decay channels.}\label{matrix-el}
\begin{tabular}{| >{\centering\arraybackslash}m{1.35cm}|  >{\centering\arraybackslash}m{5.1cm} | >{\centering\arraybackslash} m{2.6cm} |>{\centering\arraybackslash} m{2.1cm} |>{\centering\arraybackslash} m{2.25cm} |>{\centering\arraybackslash} m{2.25cm} | }
\hline
%\vspace{0.1cm}
 \multirow{3}{4em}{Decay channel} &\multicolumn{2}{c|}{} & Super-K  & \multicolumn{2}{c|}{ Sensitivities }\\
 & \multicolumn{2}{c|}{$T_{ml} $= Matrix element ($ \text{GeV}^2 $) }  &bound \cite{ParticleDataGroup:2018ovx}& \multicolumn{2}{c|}{($10^{34}$ yr)}\\
 \cline{5-6}
 &\multicolumn{2}{c|}{}&($10^{34}$ yr) & Hyper-K\cite{Hyper-Kamiokande:2018ofw}  &DUNE\cite{DUNE:2020ypp,DUNE:2015lol} \\
\hline 
\hline  
%\vspace{0.1cm}
$e^+ \,\pi^0$ &$T_{\pi^0 e^+}=\langle \pi^0|(ud)_R u_L|p\rangle_{e^+}$& $-0.131(4)(13)$&$1.6$ &$7.8$&$...$\\
$\mu^+ \, \pi^0$&$T_{\pi^0 \mu^+}=\langle \pi^0|(ud)_R u_L|p\rangle_{\mu^+}$&$ -0.118(3)(12)$&$0.77$ &$7.7$&$...$\\
$\bar{\nu}_i \, K^+ $&$T^{\prime}_{ \bar{\nu}_i K^+}=\langle K^+|(ud)_R s_L|p\rangle$& $-0.134(4)(14)$&$0.59$&$3.2$&$6.2$\\
&$T^{\prime \prime}_{\bar{\nu}_i K^+}=\langle K^+|(us)_R d_L|p\rangle$&$-0.049(2)(5)$&&&\\
$\bar{\nu}_i \, \pi^+ $&$T_{\bar{\nu}_i \pi^+ }=\langle \pi^+|(ud)_R d_L|p\rangle$& $-0.186(6)(18)$&$0.039$&$...$&$...$\\
$e^+ \, K^0$&$T_{K^0 e^+}=\langle K^0|(us)_R u_L|p\rangle_{e^+}$& $\ 0.103(3)(11)$& $0.1$ &$...$&$...$\\
$\mu^+ \, K^0$&$T_{K^0 \mu^+}=\langle K^0|(us)_R u_L|p\rangle_{\mu^+}$& $\ 0.099(2)(10)$&$0.16$&$...$&$...$\\
\hline
\end{tabular}
\end{center}
\end{table*}
and $T_{lm}$ represents hadronic matrix elements given in Table \ref{matrix-el}\,\cite{Aoki:2017puj}. Current experimental bounds from Super-K and future sensitivities of Hyper-K and DUNE on different proton decay channels  are  given in Table \ref{matrix-el}. For numerical estimates we assume $\lambda = \bar{\lambda} = \lambda' = \bar{\lambda}'$ and for convenience the mass matrix element in eqs.~(\ref{HT}) and (\ref{HbT}) is denoted by, 
\begin{eqnarray}
M_T = \sqrt{\frac{1}{(\mathcal{M}^{2}_{T^*T})^{-1}_{\  11}}} = \sqrt{\frac{1}{(\mathcal{M}^2_{\overline{T}^*\overline{T}})^{-1}_{ \ 11}}} =  \lambda^2 \frac{\upsilon^2}{M_{50}}.
\end{eqnarray}
For natural values of $\lambda$, IMDM75 predicts intermediate masses of color triplets  $M_T$.
The figures in Fig.~\ref{clpd} depict the partial lifetime of proton for charged lepton channels as a function of $\tan \beta$ ranging from $2$ to $60$.
The lower horizontal red line in each plot represents the existing Super-K bounds, while the upper horizontal red line indicates the expected sensitivities in future experiments. 
The gray shaded region in each plot represents the range of proton lifetimes that have been excluded by previous experimental searches.
%=========== neutral lepton channels ======
The decay rates for the neutral lepton channels are
\begin{widetext}
\begin{eqnarray}
\Gamma_{ \bar{\nu}_i K^+ }
&=&\mathit{F}_{K^+}A_{S_1}^2\left|\frac{g^2_5T''_{ \bar{\nu}_i K^+}}{M^2_{\chi}}\delta_{2i}
+Y^{( d)}_{11}Y^{( d)}_{ii}\left(T''_{ \bar{\nu}_i K^+ }V_{21}V^*_{1i}
+ T'_{ \bar{\nu}_i K^+}V_{11}V^*_{2i}\right)(\mathcal{M}^2_{\overline{T}^*\overline{T}})^{-1}_{\  11}\right|^2 \, ,\\
\Gamma_{ \bar{\nu}_i \pi^+} &=& \mathit{F}_{\pi^+}A_{S_1}^2T^2_{ \bar{\nu}_i \pi^+ }
\left|\frac{g^2_5}{M^2_{\chi}} +Y^{( d)\,2}_{11}\ (\mathcal{M}^2_{\overline{T}^*\overline{T}})^{-1}_{\  11}\right|^2\, .
\end{eqnarray}
\end{widetext}
\begin{figure*}[t!]\centering
\subfloat[\label{dune}$p \rightarrow \bar{\nu}_i K^+$]{\includegraphics[width=0.45\textwidth]{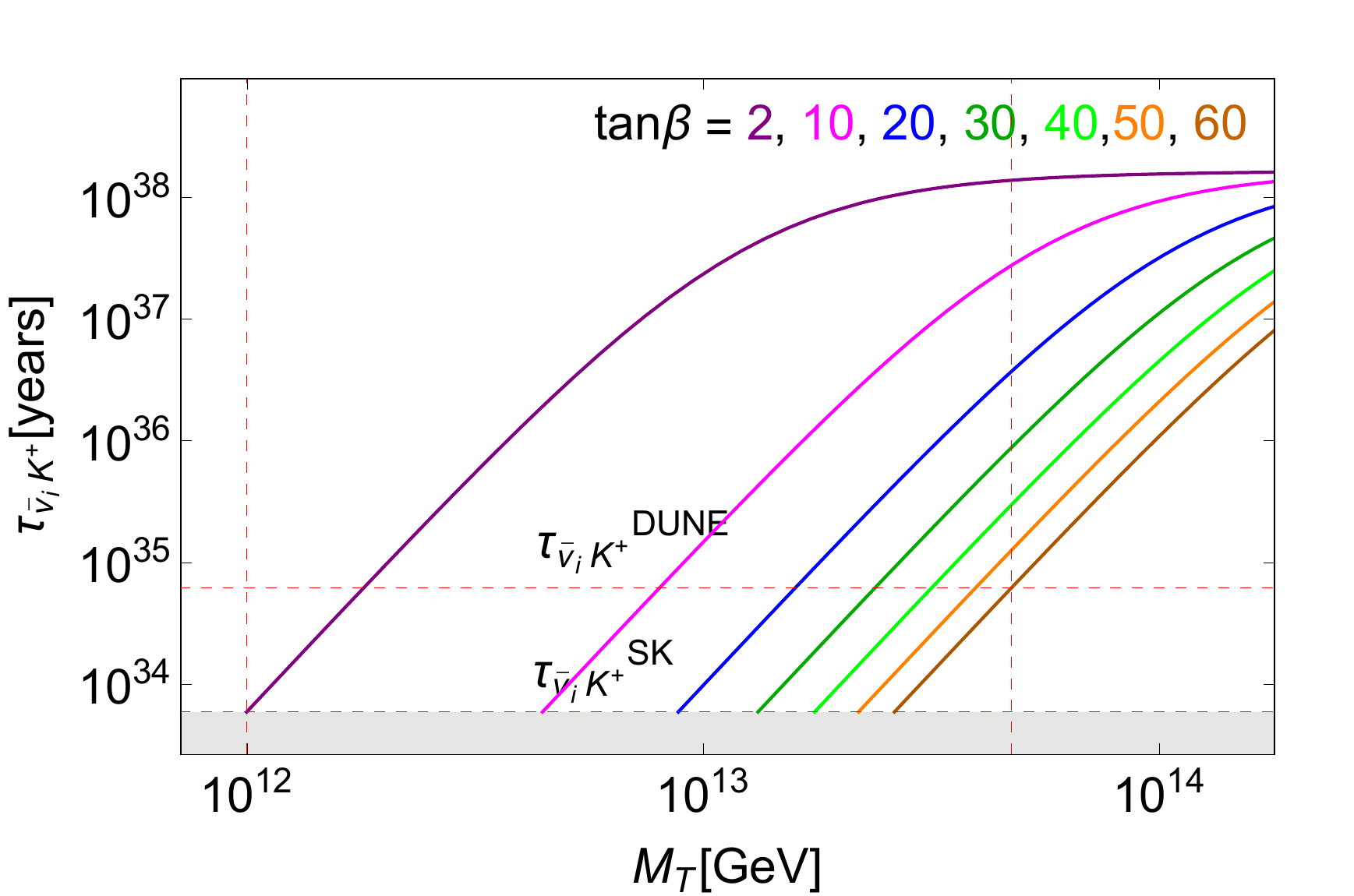}}\qquad
\subfloat[$p \rightarrow \bar{\nu}_i \pi^+$]{\includegraphics[width=0.45\textwidth]{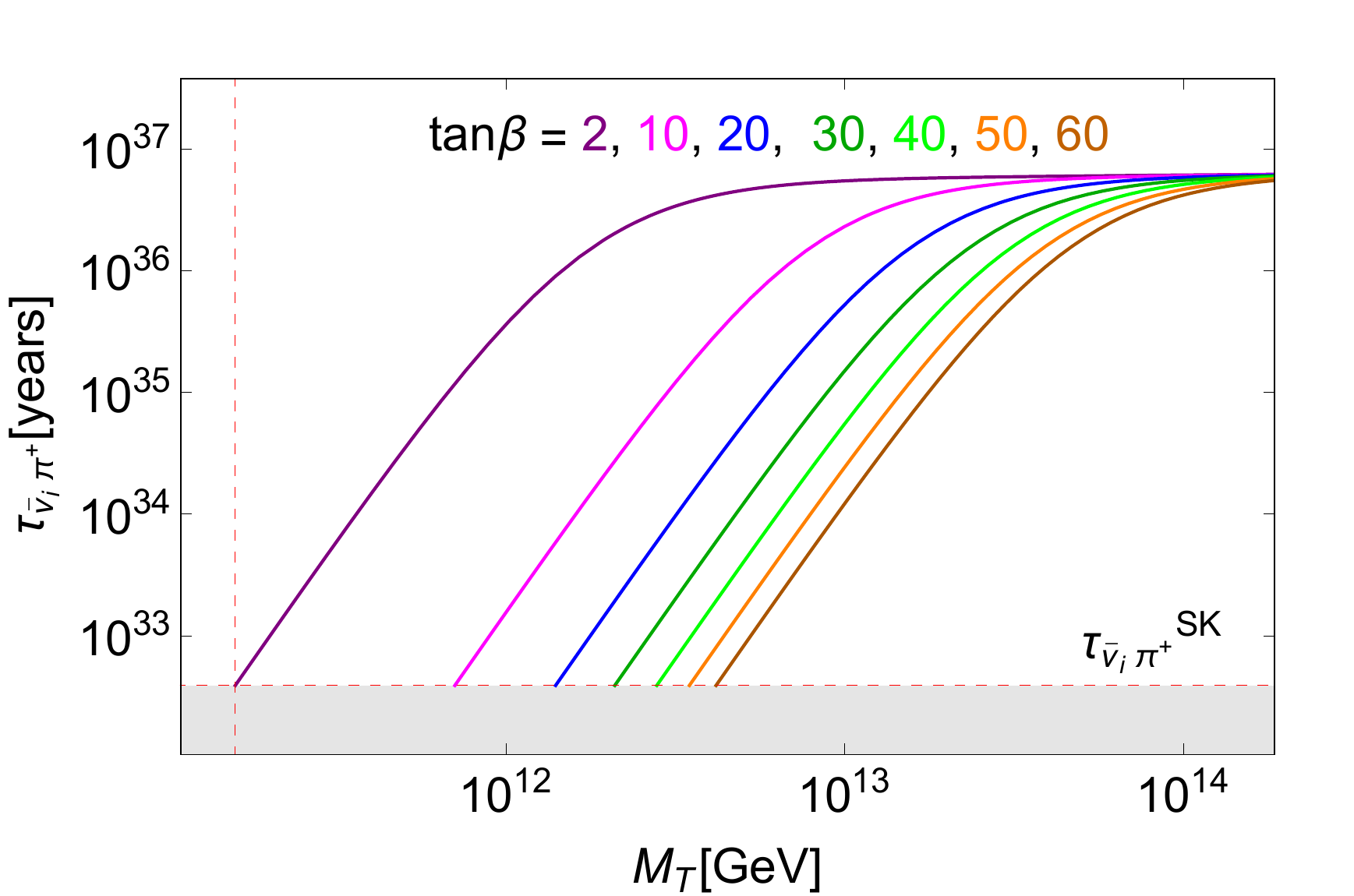}}
\caption{\label{nlpd} The partial-lifetime estimates of proton for neutral lepton decay channels as a function of $M_T$  with $\tan \beta$ in the range $2\leq\tan{\beta}\leq 60$ (increasing from left to right). The lower dashed line represents the experimental limit from Super-K and the upper dashed line in \ref{dune} represents the future DUNE limit}
\end{figure*}
Figure \ref{nlpd} displays the estimated partial-lifetime values for the neutral lepton channels. For $M_T >> 10^{14}$~GeV, the gauge boson mediation becomes dominant over the color-triplet mediation, and the partial-lifetime estimates become independent of $M_T$.
%======================================== 

\begin{figure*}[t!]\centering
\subfloat[]{\includegraphics[width=0.48\textwidth]{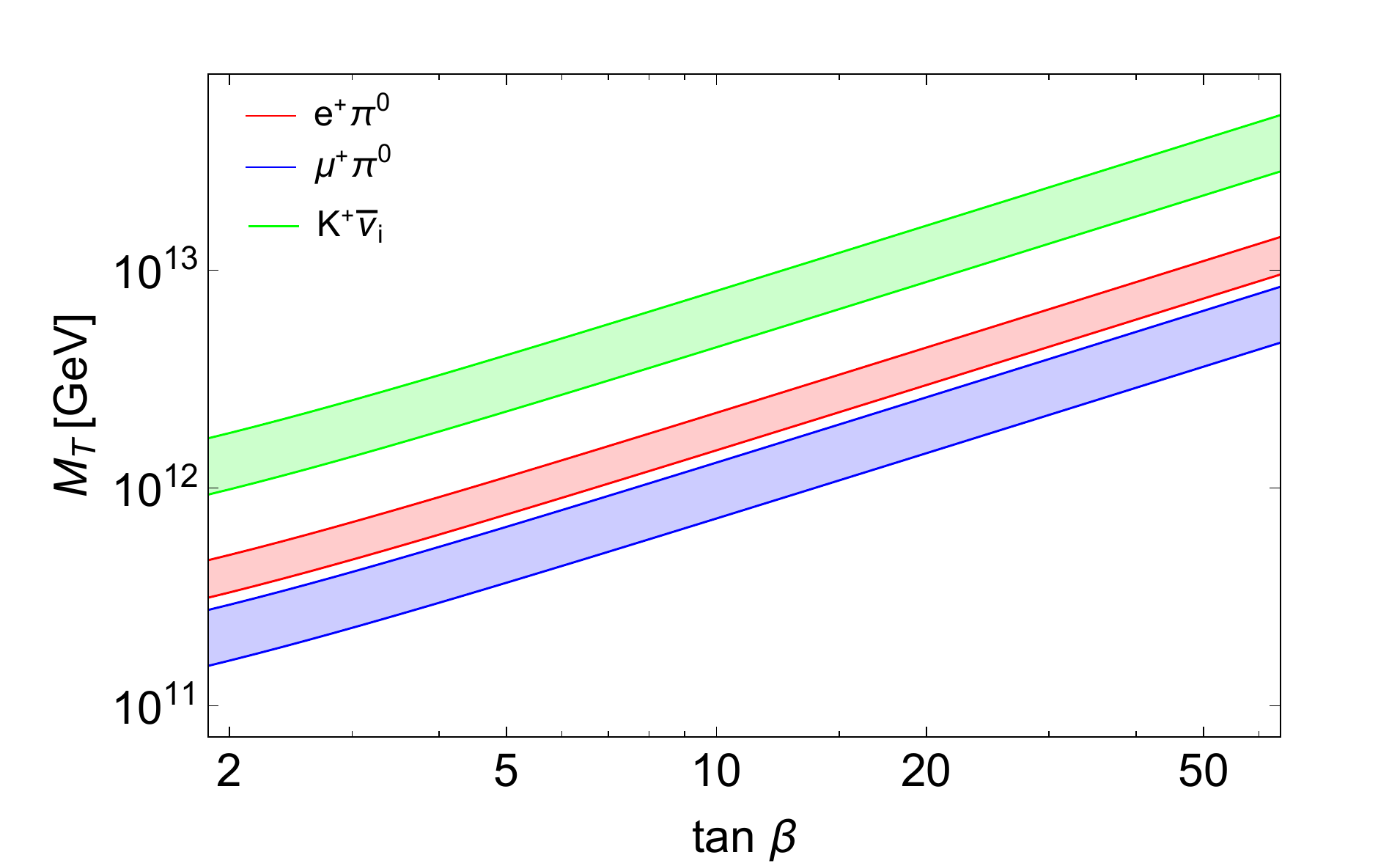}}
\subfloat[]{\includegraphics[width=0.48\textwidth]{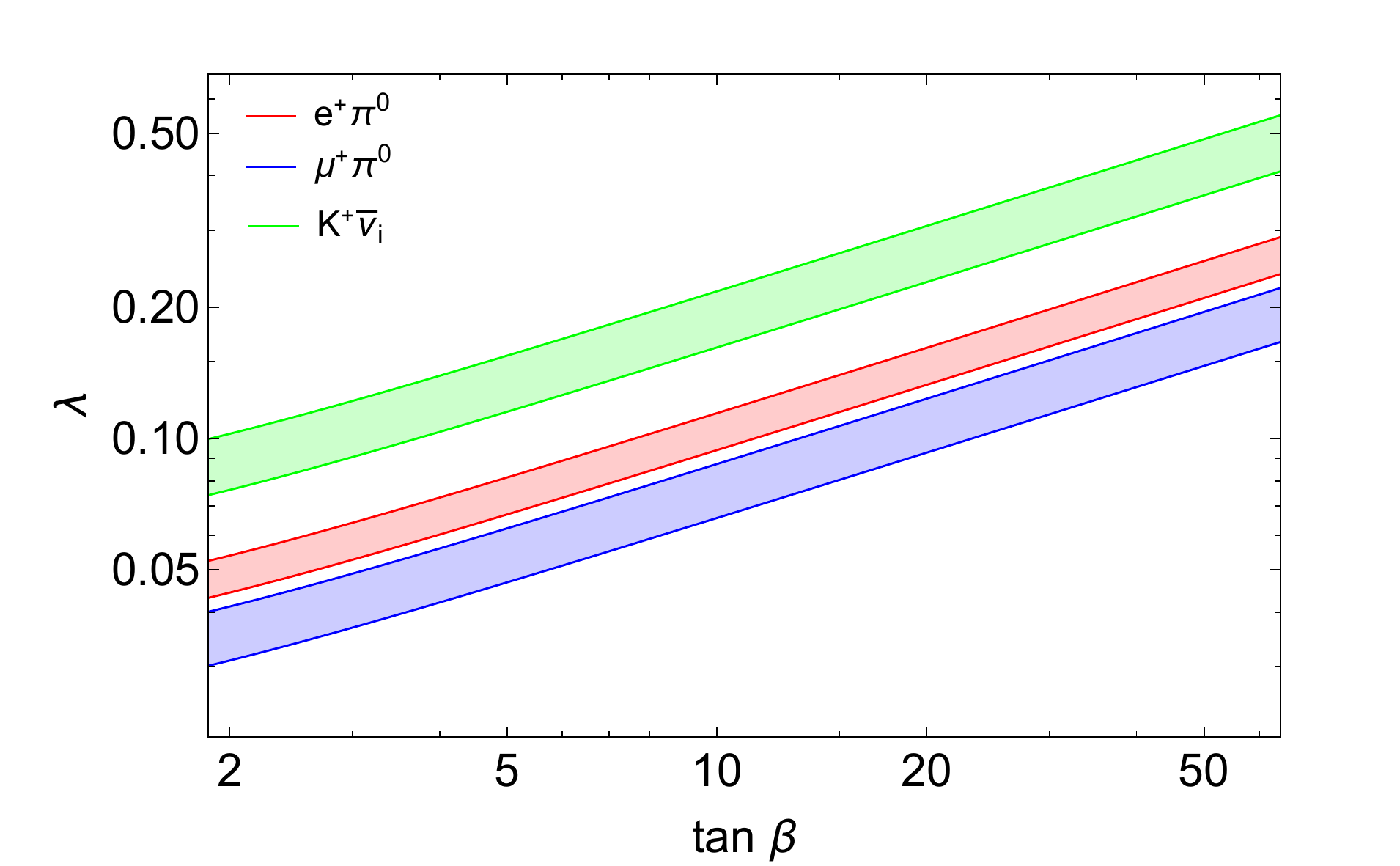}}
\caption{\label{mtlbf}  Range of $M_T$ (left panel)  and $\lambda$ (right panel) set by different proton decay channels dependent on $\tan \beta$, for observable proton decay in next-generation experiments.}
\end{figure*}

Using Super-K  bounds on proton decay channels, lower bounds on the mass of color triplets and involved couplings can be derived as
\begin{eqnarray}
 M_T &\gtrsim &\sqrt{1+\tan^2 \beta}\times M_{T}^{min} \, ,\\
\lambda 
&\gtrsim &\left(1+\tan^2 \beta\right)^{1/4}\times {\lambda_{min}} \, ,
\end{eqnarray}
where $M_T^{min}$ and $\lambda_{min}$ ,respectively, represents the lower bound on $M_T$ and $\lambda$ for a given $\tan \beta$ value and decay mode. Similarly, the observable range for $M_T$ and $\lambda$ dependent on $\tan \beta$ can be found using Hyper-K and DUNE bounds, as shown in Fig.~\ref{mtlbf}. It is worth noting that IMDM75 predicts the observable range of proton-lifetime for the next-generation experiments with somewhat natural values of the coupling, $0.03 \lesssim \lambda \lesssim 0.2$.

%=========================================

\section{IMDM75 versus other GUTs\label{vsvs}}

\begin{figure*}[ht!]\centering
\subfloat[$\Gamma_{\mu^+\pi^0}/\Gamma_{e^+\pi^0}$]{\includegraphics[width=0.5\textwidth]{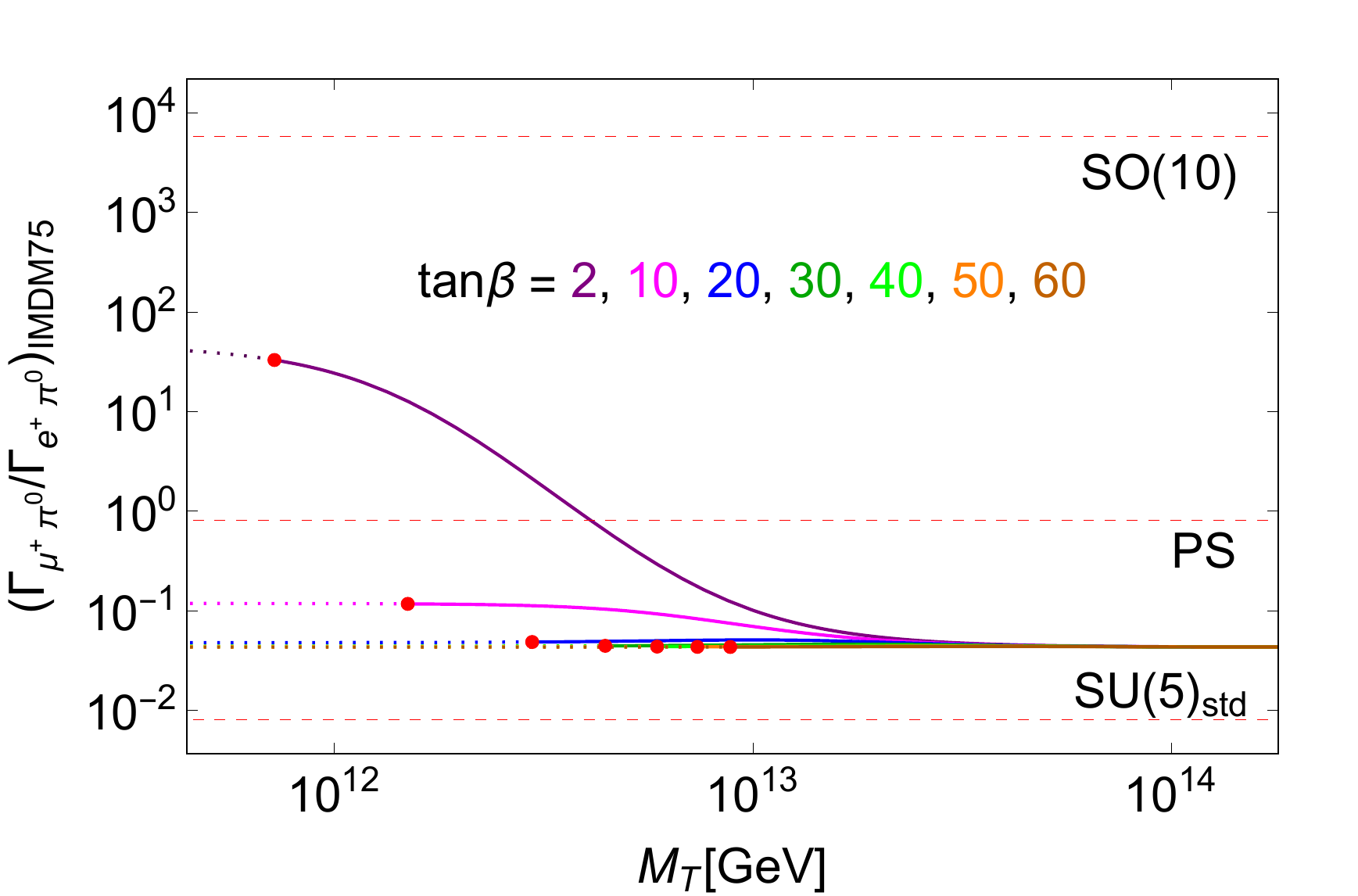}
\includegraphics[width=0.5\textwidth]{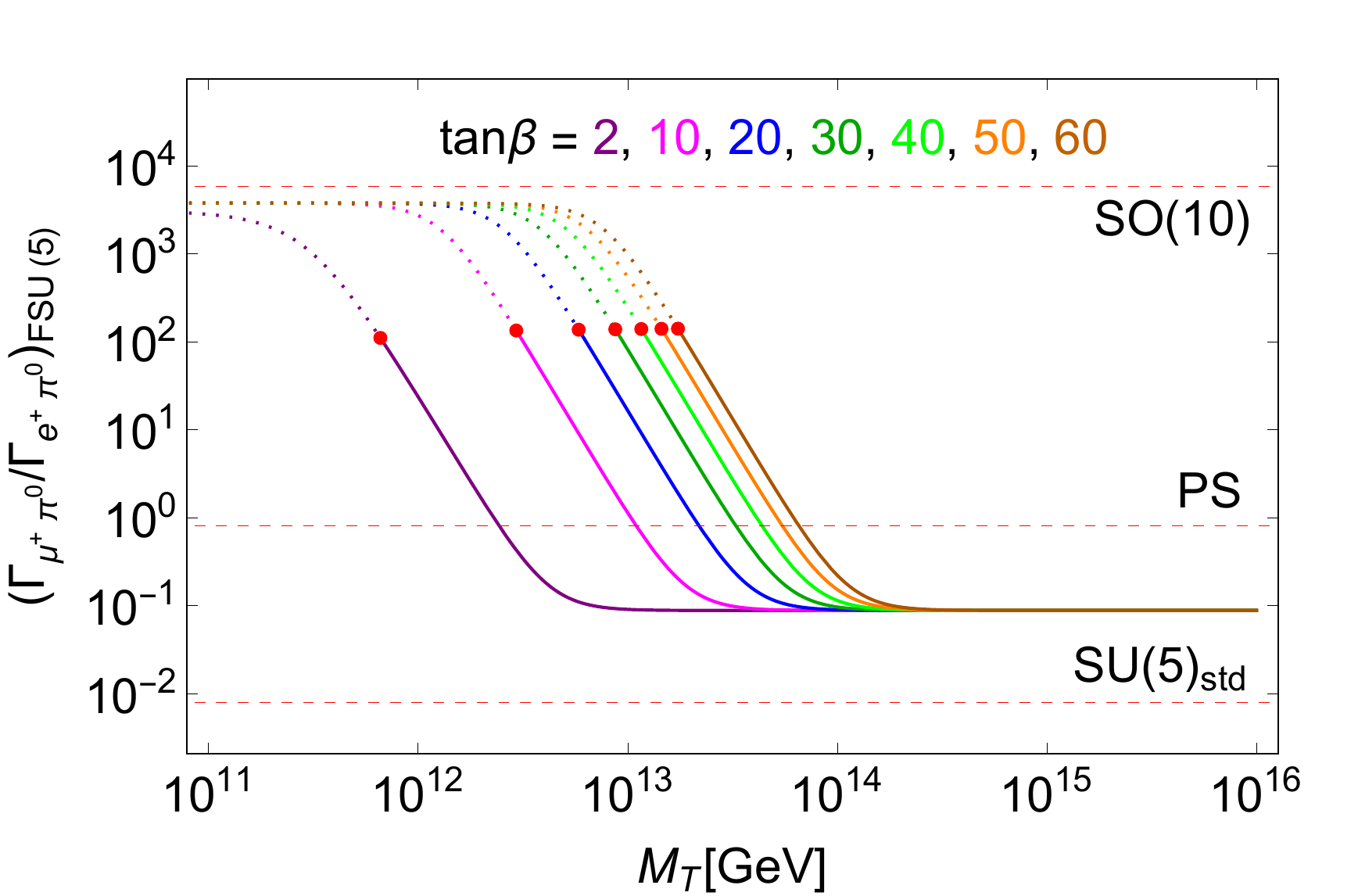}}\\
\subfloat[\label{nupiepi} $\Gamma_{\bar{\nu}_i\pi^+}/\Gamma_{e^+\pi^0}$]{\includegraphics[width=0.5\textwidth]{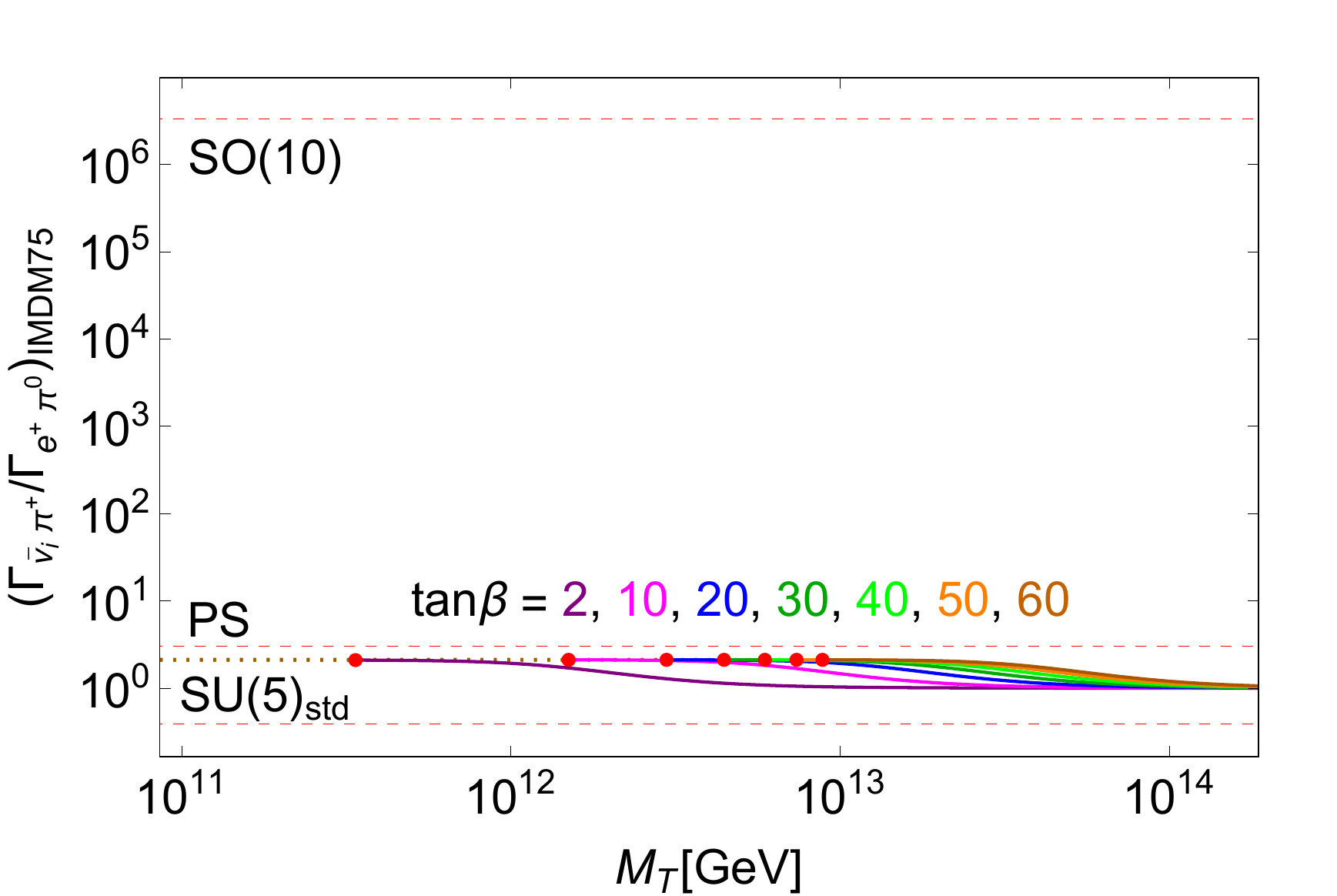}\includegraphics[width=0.5\textwidth]{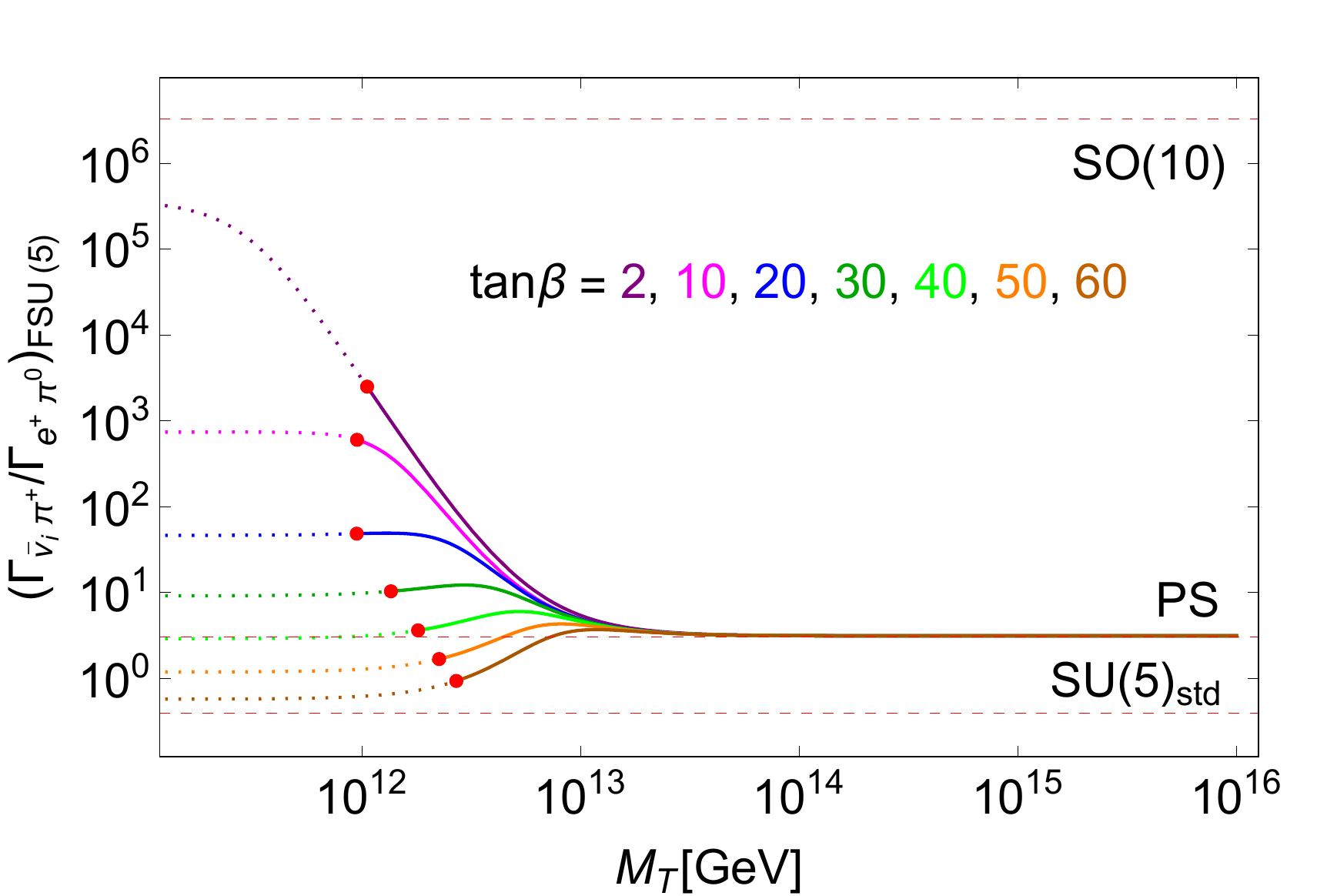}}\\
\subfloat[ \label{mukmupi} $\Gamma_{\mu^+K^0}/\Gamma_{\mu^+\pi^0}$]{\includegraphics[width=0.5\textwidth]{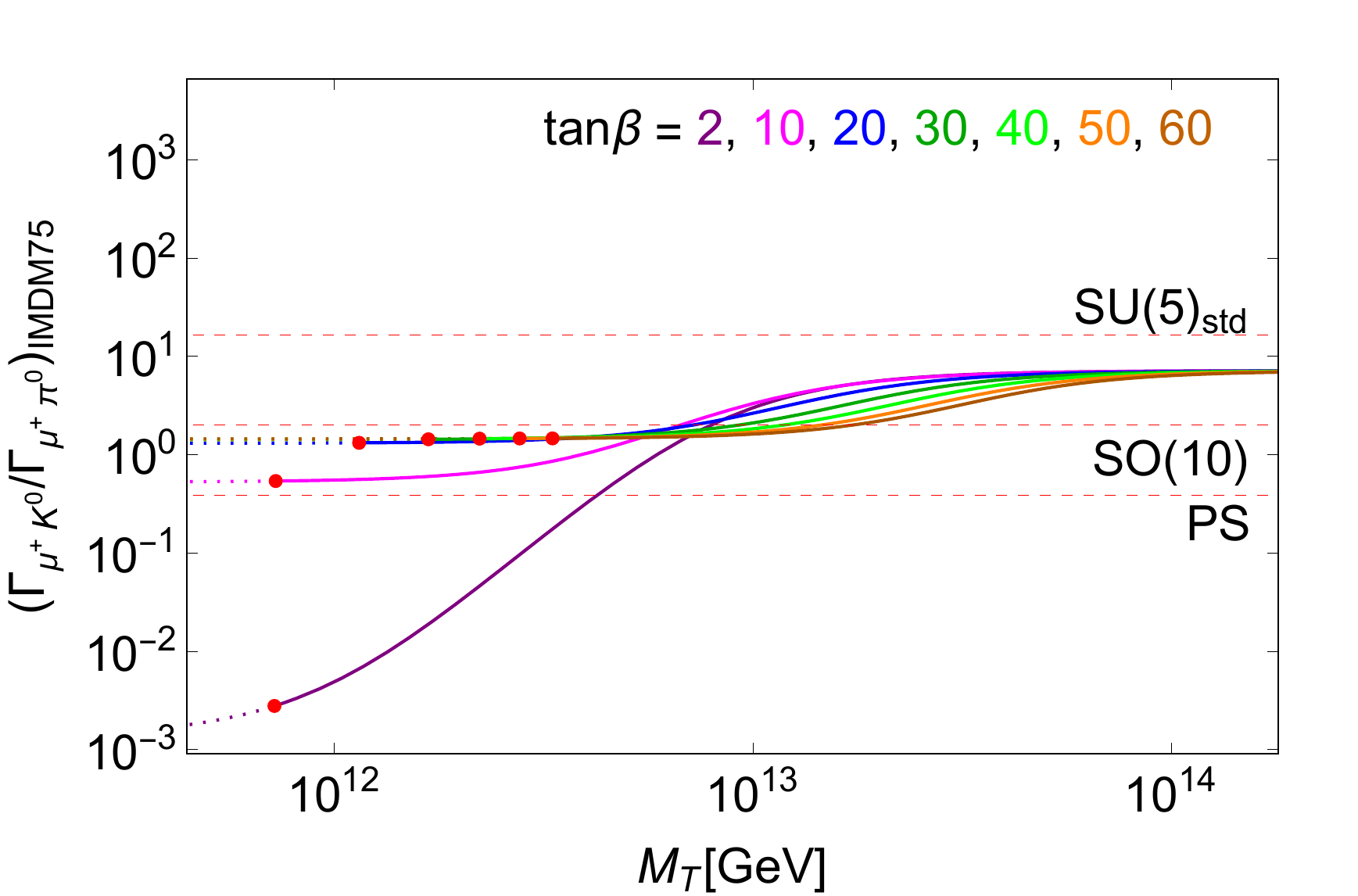}
\includegraphics[width=0.5\textwidth]{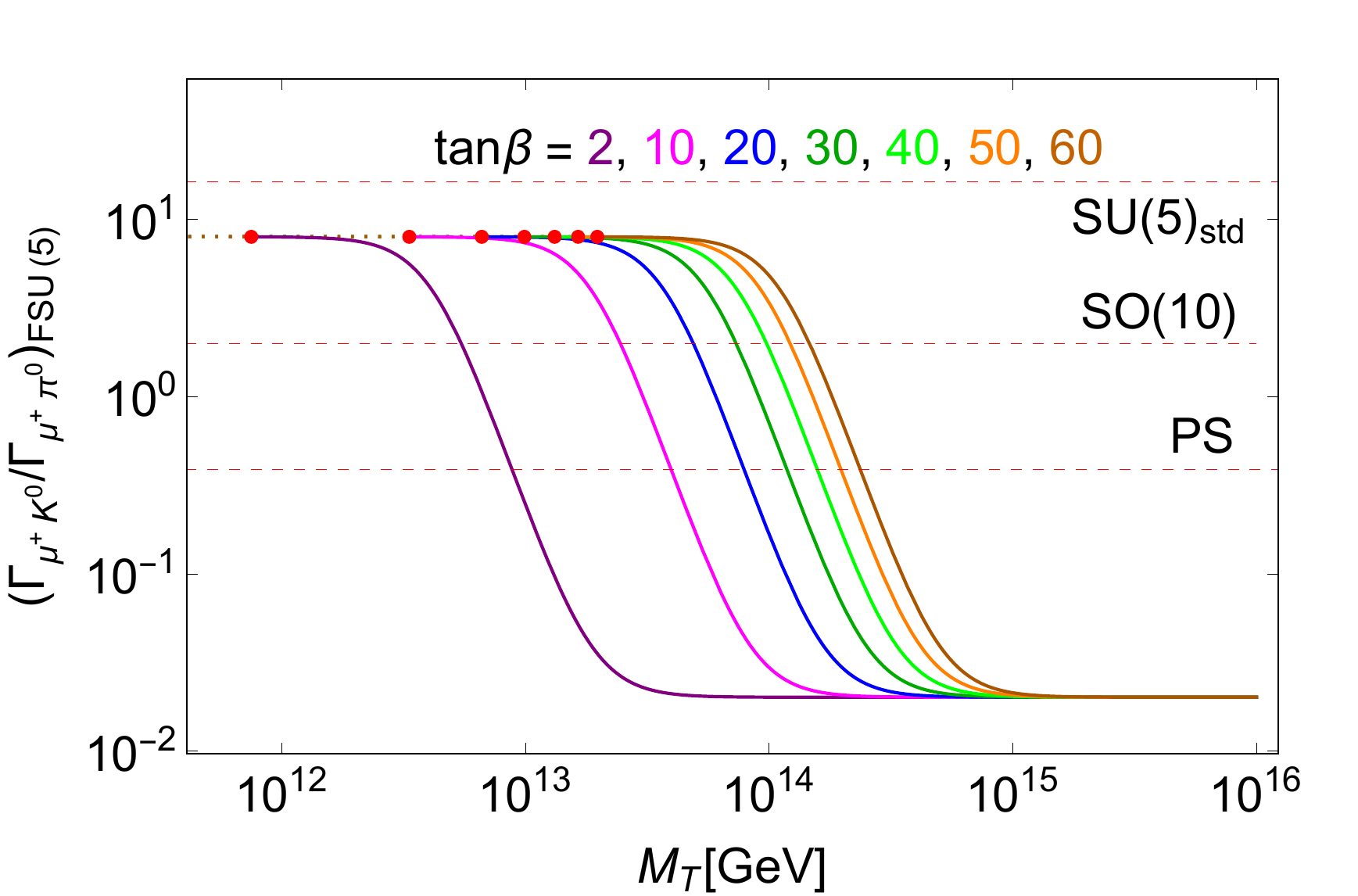}}
\caption{\label{BF1} The left panels show the estimated values of various branching fractions as a function of $M_T$ in IMDM75, where $\tan\beta$ varies from $2$ to $60$. In addition to this, the corresponding predicted values of branching fractions for $FSU(5)$ (right panels), $SU(5)_{std}$, $PS$, and $SO(10)$ (dashed red lines) from \cite{Mehmood:2020irm, Ellis:2020qad, Lazarides:2020bgy, Babu:1997js} are included for comparison. For $FSU(5)$, $M_T$ refers to the color-triplet mass defined in \cite{Mehmood:2020irm}. The solid line curves representing IMDM75 and $FSU(5)$ predictions are in agreement with the Super-K bounds, as shown by the red dots. }
\end{figure*}

\begin{figure*}[ht!]\centering
\subfloat[$\Gamma_{e^+K^0}/\Gamma_{e^+\pi^0}$]{\includegraphics[width=0.5\textwidth]{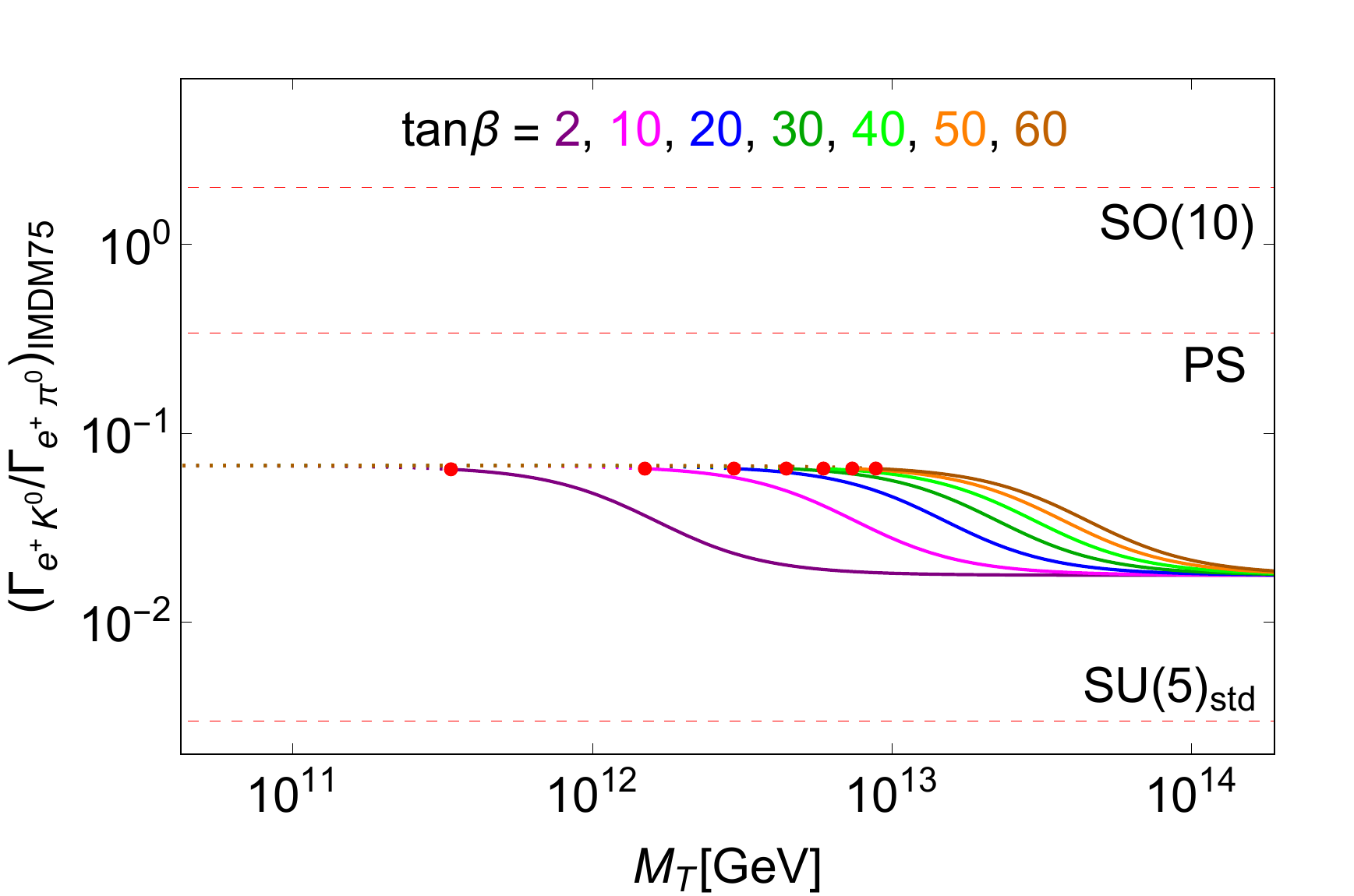}\includegraphics[width=0.5\textwidth]{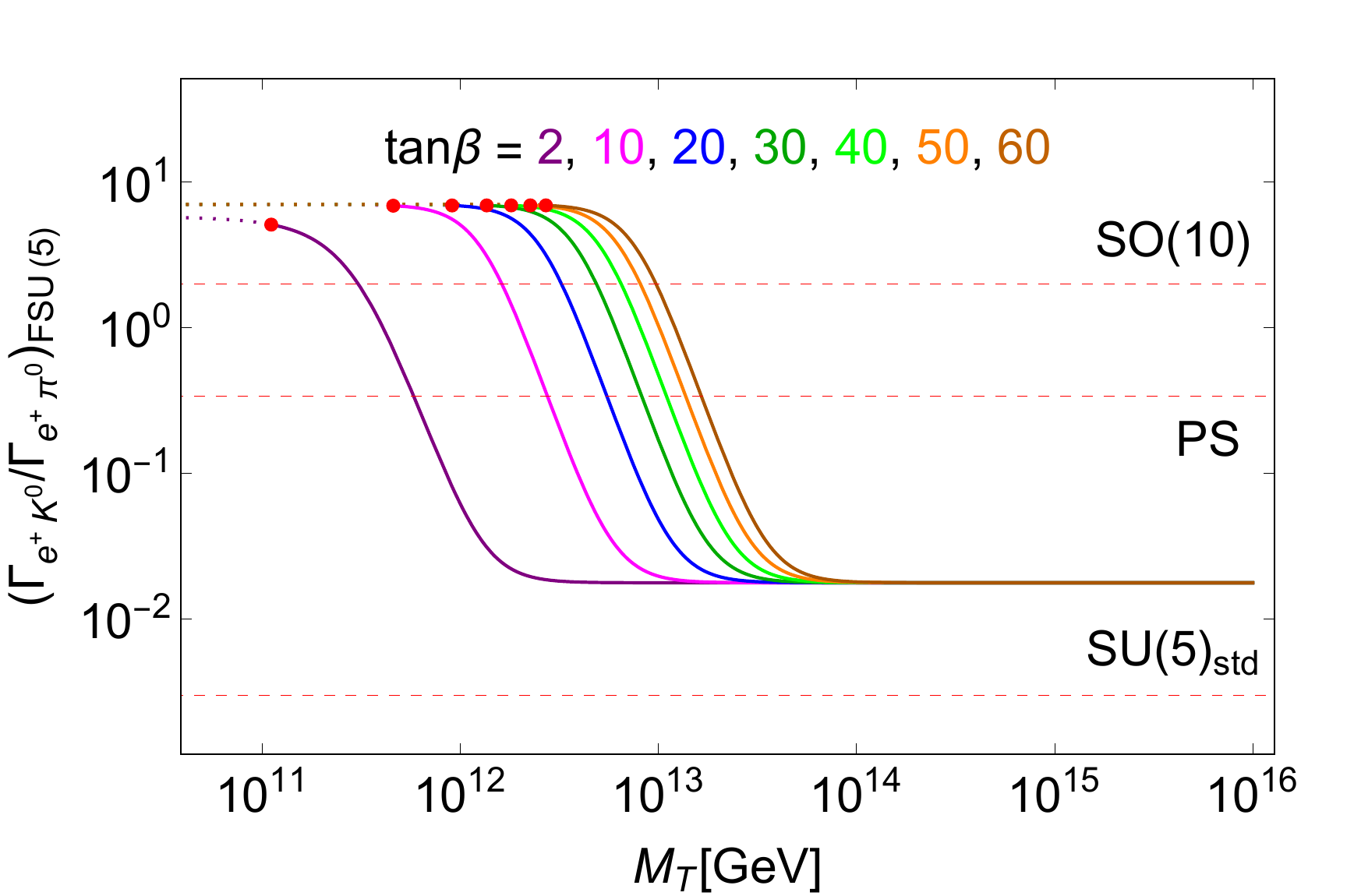}}\\
\subfloat[\label{nuknupi} $\Gamma_{\bar{\nu}_iK^+}/\Gamma_{\bar{\nu}_i\pi^+}$]{\includegraphics[width=0.5\textwidth]{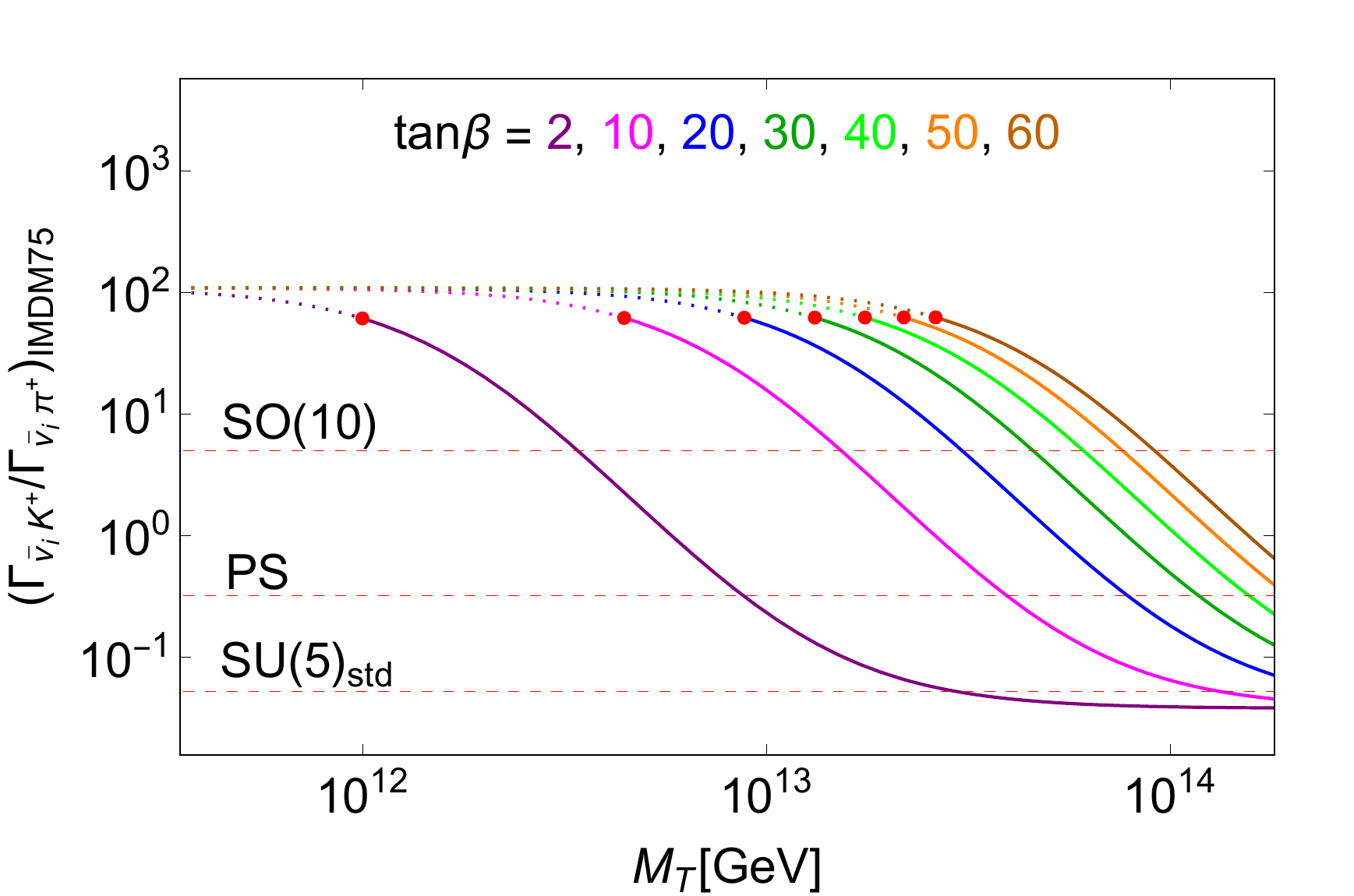} \includegraphics[width=0.5\textwidth]{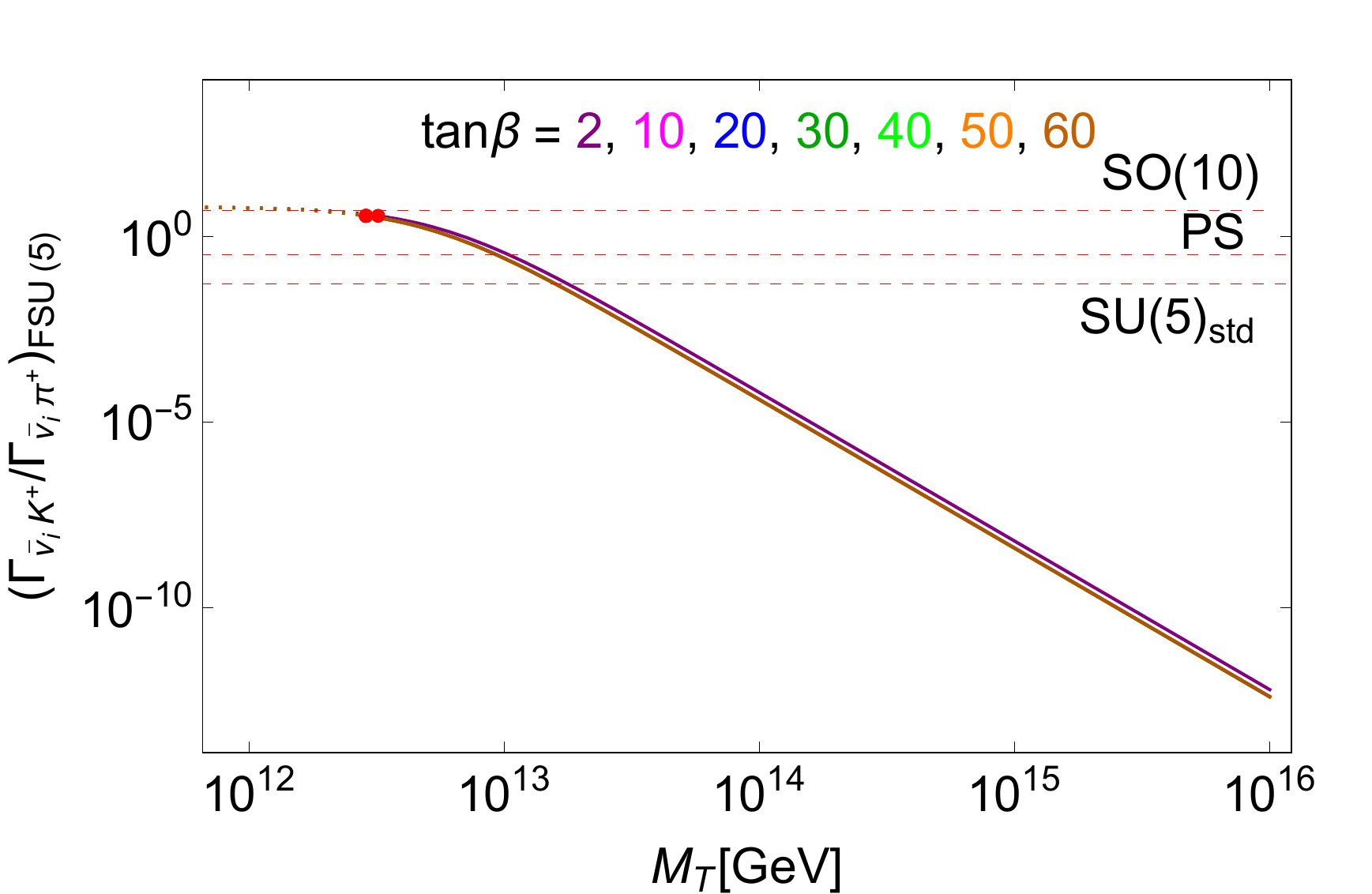}}\\
\subfloat[\label{nukek} $\Gamma_{\bar{\nu}_iK^+}/\Gamma_{e^+K^0}$]{\includegraphics[width=0.5\textwidth]{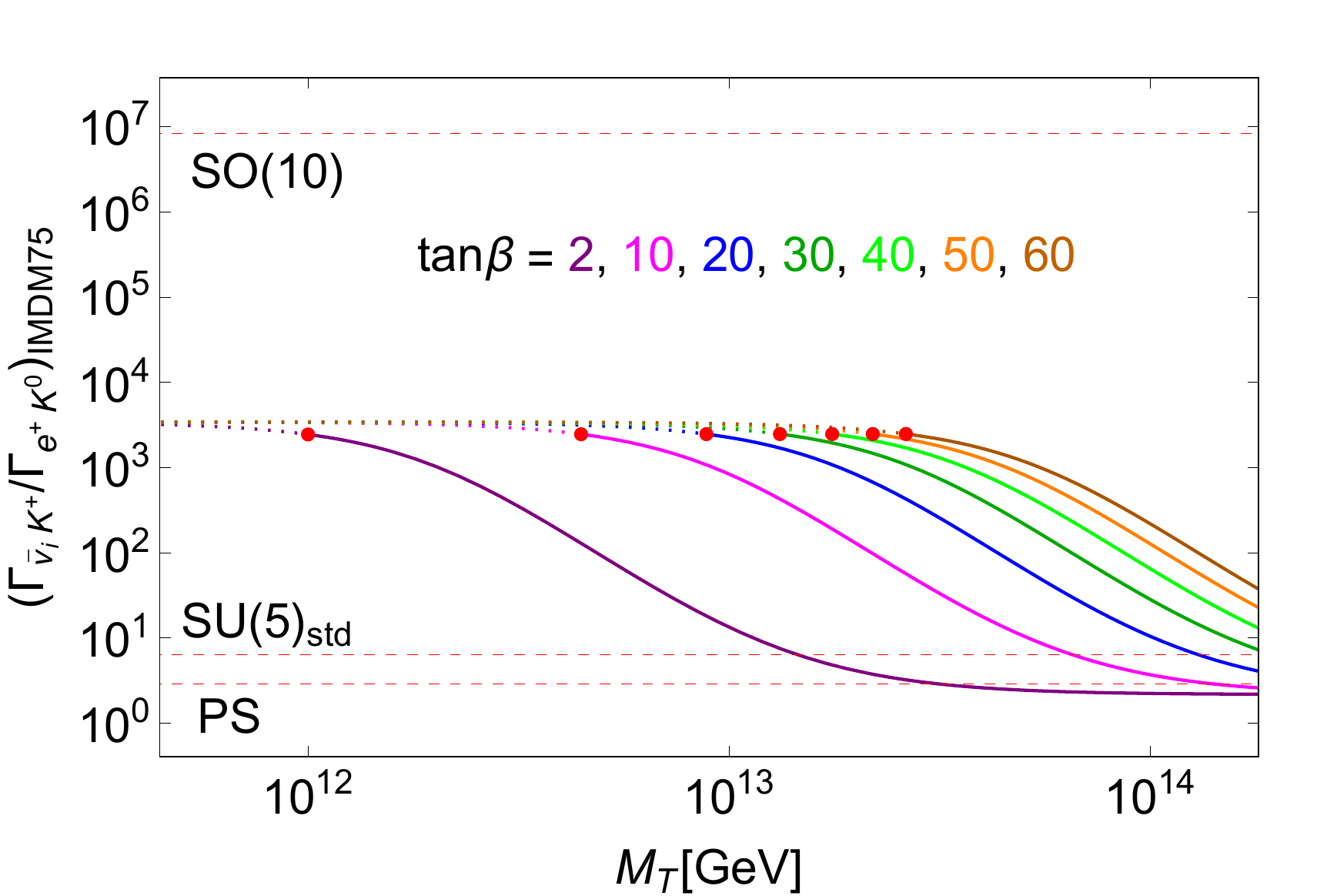}\includegraphics[width=0.5\textwidth]{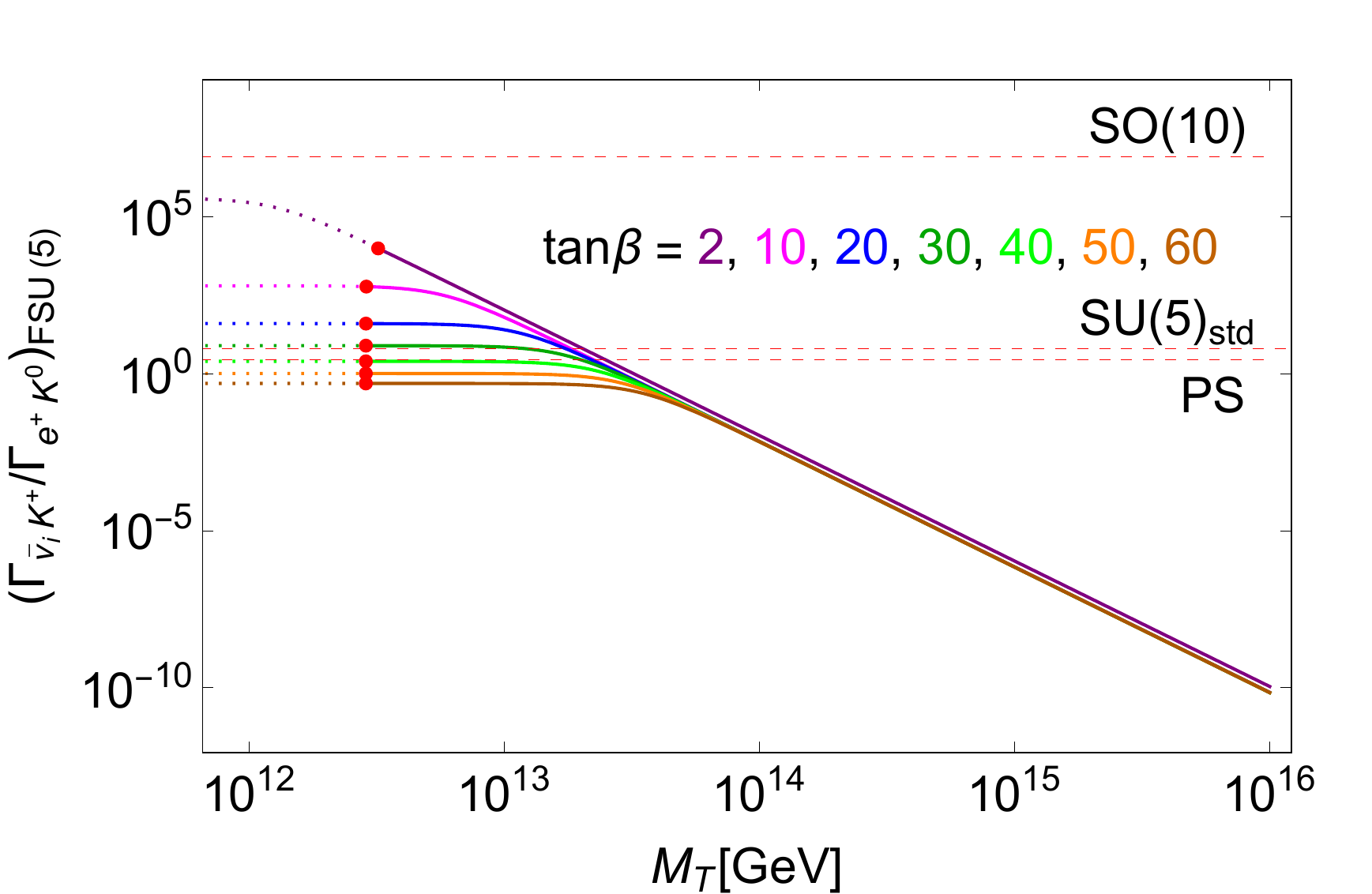}}\\
\caption{\label{BF2} The left panels show the estimated values of various branching fractions as a function of $M_T$ in IMDM75, where $\tan\beta$ varies from $2$ to $60$. In addition to this, the corresponding predicted values of branching fractions for $FSU(5)$ (right panels), $SU(5)_{std}$, $PS$, and $SO(10)$ (dashed red lines) from \cite{Mehmood:2020irm, Ellis:2020qad, Lazarides:2020bgy, Babu:1997js} are included for comparison. For $FSU(5)$, $M_T$ refers to the color-triplet mass defined in \cite{Mehmood:2020irm}. The solid line curves representing IMDM75 and $FSU(5)$ predictions are in agreement with the Super-K bounds, as shown by the red dots. 
}
\end{figure*}

To compare the proton decay predictions in IMDM75 with other GUT models, we consider the standard $SU(5)$ model with a $24$-dimensional Higgs field \cite{Ellis:2020qad} ($SU(5)_{std}$), the flipped $SU(5)$ model ($FSU(5)$) \cite{Mehmood:2020irm,Abid:2021jvn}, the Pati-Salam model ($PS$) \cite{Lazarides:2020bgy} and the Babu-Pati-Wilczek $SO(10)$ model \cite{Babu:1997js}. One can find proton decay predictions from other $SO(10)$ models in references such as \cite{Babu:1998wi,Haba:2020bls,Djouadi:2022gws}. The branching fractions for various proton decay channels are plotted in Fig.~\ref{BF1} and \ref{BF2}. The left panel displays the relationship between the branching fraction and $M_T$ in IMDM75, while the right panel presents the corresponding predictions from the flipped $SU(5)$ model.
While IMDM75 places an upper bound on the value of $M_T \lesssim v^2/M_{50}$, the flipped $SU(5)$ model allows $M_T$ to extend up to $10^{16}$ without any such constraint.

We find a significant difference in the proton lifetime behavior for the $p\rightarrow \bar{\nu} K^+$ channel compared to the predictions from the flipped $SU(5)$ model in \cite{Mehmood:2020irm}. Our plots exhibit a spread with respect to the $\tan \beta$ values, and due to the contributions from gauge boson mediation, we notice saturation at $M_T \sim 10^{15}$~GeV, which is in contrast to the flipped $SU(5)$ case where gauge boson mediation was absent and the spread in plots with respect to $\tan \beta$ values was minimal for this channel. We conclude that the $\bar{\nu}_i K^+$ channel is expected to play a crucial role in differentiating between the $SU(5)$ and flipped $SU(5)$ models.

The red horizontal lines in the figures represent the predictions of different GUT models. Specifically, the line corresponding to $SU(5)_{std}$ represents the prediction where proton decay is mainly mediated via gauge bosons. On the other hand, the lines corresponding to the Pati-Salam ($PS$) and Babu-Pati-Wilczek $SO(10)$ models represent the predictions where proton decay is dominantly mediated via color triplets. 
Thus, the IMDM75 model is found to make unique predictions for diverse branching ratios over a broad range of involved parameters, setting it apart from the three other GUT models.

\section{Gauge Coupling Unification\label{gcusec}}

In this section, we present a comprehensive two-loop analysis of gauge coupling unification in IMDM75. 
Following the breaking of the GUT symmetry, we end up with a pair of electroweak doublets with a mass of $M'_D$ and two pairs of color triplets, in addition to the minimal matter content of the MSSM. 
The color-triplet mass matrix in Eq.~(\ref{mttbar}) has eigenvalues denoted by $M_{Ti}$, where $i=1,2,3,4$.
It is worth noting that only one pair of color triplets acquires intermediate-scale masses, specifically $M_{T_1} \sim M_T$ and $M_{T_2} \sim M_T$, while the other pair, with masses $M_{T_3}$ and $M_{T_4}$, acquire GUT-scale masses.
As a result, only one pair of color triplets, with masses $M_{T_1} = M_{T_2} \sim M_T$, is pertinent for the renormalization group analysis below the GUT scale, while the other pair, with masses $M_{T_3} = M_{T_4}$, is relevant for the renormalization group analysis above the GUT scale.
We set $M_{GUT}$ as the unification scale where $g_1(GUT)=g_2(GUT)=g_{12}$, and we match IMDM75 with MSSM at $M_{GUT}$. 
The $SU(5)$ symmetry breaking into the $SU(3)_c\times SU(2)_L\times U(1)_Y$ symmetry results in the various components of $75_H$ acquiring masses (Table \ref{75m}), as given below
\begin{eqnarray}
&& M_{(1,1)} = 
 - \frac{4}{3} \lambda_{75} \upsilon  ,~ 
 M_{(3,1)} = 
 - \frac{8}{3} \lambda_{75} \upsilon ,~
 M_{(6,2)} = \frac{4}{3} \lambda_{75} \upsilon ,~   \nonumber \\
&&M_{(8,1)} = \frac{2}{3} \lambda_{75} \upsilon ,~  
  M_{(8,3)} = \frac{10}{3} \lambda_{75} \upsilon.
    \label{eq:msigma}
\end{eqnarray}
With the gauge field strength chiral superfield $\mathcal{W}$, a Planck-scale suppressed dimension-five operator can be written as
\begin{table}[t!]
\begin{center}
\caption{\label{75m} Masses and beta coefficients of Higgs and gauge boson superfield components.}
\begin{tabular}{|c|c|c|c| }
\hline
$SU(5)$ & $SU(3)_c\times SU(2)_L\times U(1)_Y$&Mass & $(b_3,b_2,b_1)$\\
\hline 
$75_H$ & $(1,1,0)$  & $M_{(1,1)}$& $(0,0,0)$\\
& $\left(3,1,5/3\right)$, $\left(\bar{3},1,-5/3\right)$  & $(3,1)$& $(1,0,10)$\\
 &$\left(3,2,-5/6\right)$, $\left(\bar{3},2,5/6\right)$  & $0$ & $(-4,-6,-10)$ \\
&  $\left(6,2,5/6\right)$, $\left(\bar{6},2,-5/6\right)$  & $M_{(6,2)}$& $(10,6,10)$\\
 &  $(8,1,0)$  & $M_{(8,1)}$& $(3,0,0)$\\
  &  $(8,3,0)$  & $M_{(8,3)}$& $(9,16,0)$\\
\hline
$5,\bar{5}$ & $(3,1,-1/3)$, $(\bar{3},1,1/3)$ &$M_{T1},M_{T2}$ &$(1,0,2/5)$\\
& $(1,2,1/2)$, $(1,2,-1/2)$ &$M'_D$ &$(0,1,3/5)$\\
\hline
$24_A$ & $(3,2,-5/6)$, $(\bar{3},2,5/6)$ &$M_{\chi}$ &$(-4,-6,-10)$ \\
\hline
\end{tabular}
\end{center}
\end{table}
\begin{eqnarray}
W_{NR} &\supset &
\frac{d}{m_P} \mathcal{W} \, \mathcal{W}\, 75_H\, .
\end{eqnarray}
The GUT scale matching conditions for the gauge couplings depend on the coupling $d$ of the above dimension-five operator \cite{Hisano:1997nu,Huitu:1999eh,Ellis:2021fhb,Evans:2021hyx} and on the masses of order GUT scale or lower  as,
\begin{eqnarray}
&&\frac{1}{g_1^2(Q)}=\frac{1}{g_5^2(Q)}+\frac{1}{8\pi^2} \biggl[
 10\ln \biggl( \frac{Q}{M_{(3,1)}} \biggr)
+ 10\ln \biggl(\frac{Q}{M_{(6,2)}}\biggr) 
\nonumber \\
&& +\frac{2}{5}\ln\left(\frac{Q}{M_{T1}}\right)
    +\frac{2}{5}\ln\left(\frac{Q}{M_{T2}}\right)
    +\frac{3}{5}\ln\left(\frac{Q}{M'_{D}}\right)\nonumber \\
&&    -10\ln \left(\frac{Q}{M_{\chi}}\right)
\biggr]+\frac{5}{2}\left(\frac{8d\,\upsilon}{m_P}\right)~,\\
&&\frac{1}{g_2^2(Q)}
=\frac{1}{g_5^2(Q)}+\frac{1}{8\pi^2} 
\biggl[ 
6\ln \biggl(\frac{Q}{M_{(6,2)}}\biggr) 
+ 16\ln \biggl(\frac{Q}{M_{(8,3)}}\biggr)\nonumber \\
&&-6 \ln\left(\frac{Q}{M_{\chi}}\right)
+\ln\left(\frac{Q}{M'_D}\right)
\biggr]-\frac{3}{2}\left(\frac{8d\,\upsilon}{m_P}\right) ~,\\
&&\frac{1}{g_3^2(Q)}=
\frac{1}{g_5^2(Q)}+\frac{1}{8\pi^2}
\biggl[
10\ln \biggl( \frac{Q}{M_{(6,2)}} \biggr)
+\ln \biggl( \frac{Q}{M_{(3,1)}} \biggr)\nonumber \\
&&+ 3\ln \biggl( \frac{Q}{M_{(8,1)}} \biggr) 
+ 9\ln \biggl( \frac{Q}{M_{(8,3)}}\biggr)
+\ln\left(\frac{Q}{M_{T1}}\right) \nonumber \\
&&+ \ln\left(\frac{Q}{M_{T2}}\right)-4\ln\left(\frac{Q}{M_{\chi}}\right)
\biggr]-\frac{1}{2}\left(\frac{8d\,\upsilon}{m_P}\right)~.
\end{eqnarray}
By eliminating $d$, we arrive at the following constraint,
\begin{eqnarray}\label{mgmgut}
\frac{6}{g_2^2(Q)}-\frac{8}{g_3^2(Q)}+\frac{2}{g_1^2(Q)}=
\frac {27}{5{\pi}^{2}}\ln  \left( 5^{-\frac{5}{9}}
\frac{M_G}{Q} \right),
\end{eqnarray}
where, 
\begin{eqnarray}
M_G &=&
\left(\frac{M_{(8,1)}^5M_{\chi }^{10}M_{\text{T1}}^3M_{\text{T2}}^3}{ M_D^{\prime \,3}}\right)^{ {1}/{18}}\, .
\end{eqnarray}
\begin{figure}[t!]\centering
\includegraphics[width=0.5\textwidth]{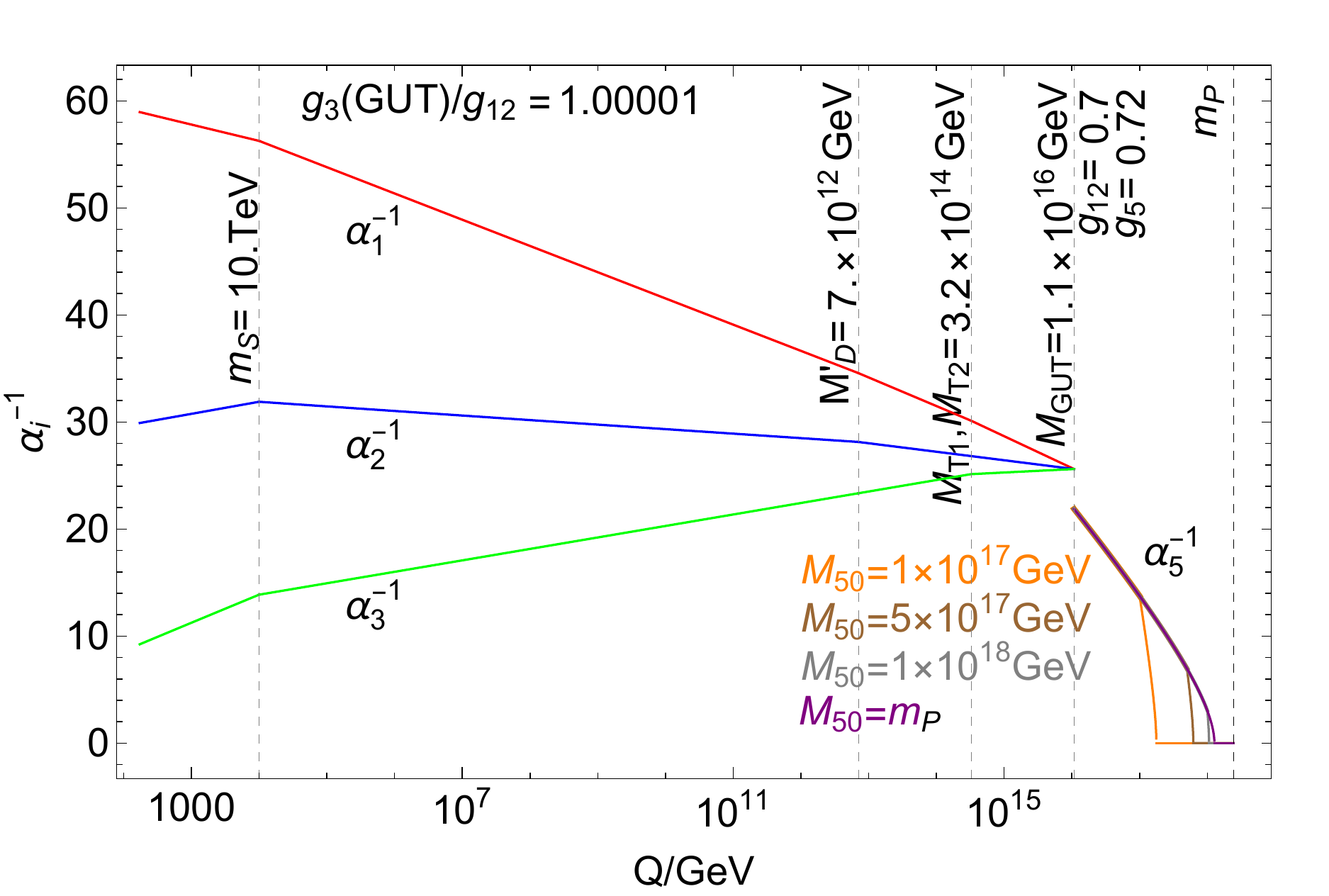}
\caption{\label{gcu} The gauge coupling unification with two loop RGEs for IMDM75 and  MSSM plus one pair of electroweak doublet and color-triplet at an intermediate scale. We assumed all supersymmetric particles of degenerate mass $m_S$. We achieved gauge coupling unification with the GUT scale VEV $\upsilon = 5.8 \times 10^{15}$ GeV.}
\end{figure}
To investigate gauge coupling unification, we employ two-loop  renormalization group equations (RGEs) for MSSM, with an extra pair of the electroweak doublet and the color-triplet at the intermediate scale,  as depicted in Fig.~\ref{gcu}. This analysis yields $M_{GUT} = 1.1 \times 10^{16}$~GeV,
$g_{12} = 0.7$, and $g_3(GUT)/g_{12} = 1.00001$. At the GUT scale, applying Eq.~(\ref{mgmgut}) leads to the following constraint
\begin{eqnarray}
M_G &= & 5^{5/9}e^{{0.00029}/{g_{12}^2}} M_{GUT}\, .
\end{eqnarray}
Assuming $M_{T1}=M_{T2}\sim 10^2 \times M_{T3}=M_{T4}$ and $\lambda \sim 1$ we obtain the following expression
\begin{eqnarray}
\upsilon &= & 4.7 \times 10^{15} \ \text{GeV} \times g_5^{-2/3}\, .
\end{eqnarray}
To find the matching condition for $g_5$ we employ the following constraint
\begin{eqnarray}
\frac{5}{g_1^2}+\frac{3}{g_2^2}-\frac{2}{g_3^2}
=
\frac{72\, d\, \upsilon}{m_P}+\frac{6}{g_5^2}+\frac{33 }{4 \pi ^2}\ln \biggl(\frac{Q}{M_{g_5}}\biggr)\, ,
\end{eqnarray}
with
\begin{eqnarray}
M_{g_5}= \biggl(\frac{M_{(3,1)}^8 M_{(6,2)}^8M_{(8,3)}{}^5 M'_D}{M_{(8,1)} M_{\chi }^{10} }\biggr)^{1/11}\, .
\end{eqnarray}
When evaluated at the GUT scale, the aforementioned constraint yields the following expression
\begin{eqnarray}
45.9\, g_5^{2/3}\biggl[5.4- {2.9\,}{g_5^{-2}}-0.6\, \ln ( g_5 )\biggr]=d.
\end{eqnarray}
In Fig.~\ref{gcu}, we depict the behavior of the gauge coupling $g_5$ above the GUT scale for a value of $d\sim 0.2$, which is adequate to guarantee perturbative gauge coupling \cite{Ellis:2021fhb}. The presence of GUT multiplets with large representations leads to $g_5$ becoming nonperturbative just above the mass scale $M_{50} \sim 10^{17}-10^{18}$~GeV.
\section{\label{con}Conclusion}

The article investigates proton decay with partial lifetimes in IMDM75, improved missing doublet $SU(5)$ model, that can be tested in the next-generation experiments.
The model incorporates an anomalous $U(1)_A$ symmetry to prevent the occurrence of the $5_h\bar{5}_h$ term at all orders. 
For GUT symmetry breaking and implementation of the missing partner mechanism, a $75$-plet GUT Higgs field ($75_H$) is utilized in the presence of two pairs of $5$ and $50$-plet fields. 
To prevent fast proton decay and achieve successful GCU in the improved model, the presence of a second pair of $5$ and $50$-plet fields is necessary. This allows an extra pair of electroweak doublets and one pair of color triplets to acquire masses naturally at the intermediate scale.
The model predicts observable proton decay via the naturally obtained intermediate masses of chirality nonflipping mediation of color triplets, while the chirality flipping dimension-five proton decay mediated via color triplets is sufficiently suppressed to avoid fast proton decay. 
The model's predictions for various branching ratios are compared with those of $FSU(5)$, $SU(5)_{std}$, PS model, and $SO(10)$, and in this regard the significance of the $\bar{\nu}_i K^+$ channel is emphasized. 
The model presented in this article provides unique predictions for different branching ratios over a broad range of involved parameters.
Finally, the matching conditions for the gauge couplings involved are provided, and the compatibility of gauge coupling unification with observable proton decay lifetimes is demonstrated.

%%%%%%%%%%%%%%%%%%%%%%%%%%%%%%%%%%%%%%%%%%%%%%%%%%%%%%%
\begin{acknowledgments}
M. M. thanks Zurab Tavartkiladze for useful discussion. 
\end{acknowledgments}

\bibliographystyle{unsrt}

\bibliography{draft-SU(5)MDM-PRD}{}% Produces the bibliography via BibTeX.

\end{document}